\documentclass[a4paper,prd,aps,11pt,nofootinbib,notitlepage,superscriptaddress]{revtex4-1}
\usepackage[english]{babel}
\usepackage[utf8]{inputenc}	% Allows for writing special charachters in the tex-file

\usepackage{amsmath,amssymb,amsfonts,dsfont} 	% Standard mathematics
\usepackage{xspace}
\usepackage{hyperref}
\usepackage{multirow}

\usepackage{graphicx}
\graphicspath{{./Figures/}}
\renewcommand{\arraystretch}{1.2}
\usepackage{siunitx}
\usepackage{color}	% Colors
\definecolor{rossoCP3}{cmyk}{0,.88,.77,.40}
\definecolor{blaa}{rgb}{0.2,0.2,0.6}

\hypersetup{
	colorlinks,
	bookmarksopen,
	bookmarksnumbered,
	citecolor=blaa, 		%color of links to bibliography
	linkcolor=rossoCP3,	%color of internal links
	urlcolor=rossoCP3,			%color of external links
}

\usepackage[noindentafter]{titlesec}

\newcommand{\brakets}[1]{\left\langle #1 \right\rangle}

\newcommand{\Tr}{\mathrm{Tr}}
\newcommand{\tr}{\text{tr}}
\newcommand{\transpose}{^{\mathrm{T}}}
\newcommand{\hc}{\; + \; \mathrm{h.c.} \;}
\newcommand{\andeq}{\quad \mathrm{and} \quad}

\newcommand{\LL}{\mathrm{L}}

\newcommand{\TC}{\mathrm{TC}}
\newcommand{\EW}{\mathrm{EW}}
\newcommand{\SM}{\mathrm{SM}}

\newcommand{\SU}{\mathrm{SU}}
\newcommand{\Sp}{\mathrm{Sp}}

\newcommand{\tcf}{\mathcal{F}}
\newcommand{\tcs}{\mathcal{S}}

\newcommand{\Fupdn}{\tcf_\updownarrow}
\newcommand{\Fupbar}{\bar{\tcf}_\uparrow}
\newcommand{\Fdnbar}{\bar{\tcf}_\downarrow}

\newcommand{\ubar}{{\bar{u}}}
\newcommand{\dbar}{{\bar{d}}}
\newcommand{\ebar}{{\bar{e}}}
\newcommand{\nubar}{{\bar{\nu}}}

\newcommand{\spur}[2]{\psi^{#1}\phantom{}_{#2} }
\newcommand{\spurbar}[2]{{\psi^\dagger}^{#1 #2}}

\begin{document}
%%%%%%%%%%%%%%TITLE AFFILIATIONS ETC%%%%%%%%%%%%%%%%%%%%%%%%%%%%%%%%%%%%%%%%%%%%%%%%%%%%%%%%%%%%

\title{\texorpdfstring{\LARGE\color{rossoCP3}  Flavor Physics and Flavor Anomalies\\in Minimal Fundamental Partial Compositeness}{Flavor Physics in Minimal Fundamental Partial Compositeness}}
\author{Francesco {\sc Sannino}}
\email{sannino@cp3.dias.sdu.dk}
\affiliation{CP$^3$-Origins, University of Southern Denmark, Campusvej 55, 5230 Odense, Denmark}
\affiliation{Danish IAS, University of Southern Denmark,  Odense, Denmark}
%\affiliation{CERN, Theory Division, Geneva, Switzerland.}
\author{Peter {\sc Stangl}}
\email{peter.stangl@tum.de}
\affiliation{Excellence Cluster Universe, TUM, Boltzmannstr.~2, 85748~Garching,
Germany}
\author{David M. {\sc Straub}}
\email{david.straub@tum.de}
\affiliation{Excellence Cluster Universe, TUM, Boltzmannstr.~2, 85748~Garching,
Germany}
\author{Anders Eller {\sc Thomsen}}
\email{aethomsen@cp3.sdu.dk}
\affiliation{CP$^3$-Origins, University of Southern Denmark, Campusvej 55, 5230 Odense, Denmark}
%%%%%%%%%%%%%%%%%%%%%%%%%%%%%%%%%%%%%%%%%%%%%%%%%%%%%%%%%%%%%%%%%%%%%%%%%%

\begin{abstract}\noindent
Partial compositeness is a key ingredient of models where the electroweak symmetry is broken by a composite Higgs state. Recently, a UV completion of partial compositeness was proposed, featuring a new strongly coupled gauge interaction as well as new fundamental fermions and scalars. We work out the full flavor structure of the minimal realization of this idea and investigate in detail the consequences for flavor physics. While CP violation in kaon mixing represents a significant constraint on the model, we find many viable parameter points passing all precision tests. We also demonstrate that the recently observed hints for a violation of lepton flavor universality in $B\to K^{(*)}\ell\ell$ decays can be accommodated by the model, while the anomalies in $B\to D^{(*)}\tau\nu$ cannot be explained while satisfying LEP constraints on $Z$ couplings.
\\[.3cm]
{\footnotesize  \it Preprint: CP$^3$-Origins-2017-058 DNRF90}
\end{abstract}

\maketitle

\section{Introduction}\label{sec:intro}
New composite dynamics is a long standing framework for electroweak (EW) symmetry breaking, providing a promising solution to the hierarchy problem by removing the Higgs boson as an elementary scalar. Rather than the Higgs boson gaining a vacuum expectation value, the breaking of the EW symmetry is instead brought on by the formation of a condensate in a new, strongly interacting sector of the theory. In modern composite models the Higgs boson is realized as a pseudo Nambu-Goldstone Boson (pNGB) keeping it light compared to the scale of the new dynamics~\cite{Kaplan:1983fs}.

A major challenge in constructing a successful model of strong EW symmetry breaking is providing masses to the Standard Model (SM) fermions.
In this respect, the idea of partial compositeness has proved popular~\cite{Kaplan:1991dc}; here the SM fermions mix with composite fermions of appropriate quantum numbers to gain their masses.
Most of the phenomenological studies of composite Higgs models have focused on simplified models implementing the partial compositeness mechanism at low energies, without specifying the UV completion. Constructing an explicit UV completion is important not only to lend credibility to the partial compositeness framework in general, but also since it may lead to specific correlations that can be tested in low-energy precision experiments.
In a recent development, Fundamental Partial Compositeness (FPC) models were proposed%
\footnote{%
Other almost UV completions of partial compositeness have also been proposed in the literature. See \cite{Caracciolo:2012je} for supersymmetric constructions and \cite{Barnard:2013zea,Ferretti:2013kya,Ferretti:2014qta,Vecchi:2015fma} for purely fermionic constructions. It remains to be seen whether the required large anomalous dimensions can be achieved in the purely fermionic constructions \cite{Pica:2016rmv}.}
that feature new fermions and scalars charged under a strong ``technicolor'' (TC) force.
In these models, the SM fermions gain masses as a result of fundamental Yukawa interactions between SM fermions, TC fermions, and TC scalars~\cite{Sannino:2016sfx}.
This allowed for a controlled construction of the complete effective field theory (EFT) respecting all the symmetries of the Minimal FPC (MFPC) model \cite{Cacciapaglia:2017cdi}. First principle lattice simulations have begun to investigate this novel dynamics in \cite{Hansen:2017mrt} while the pioneering work without techniscalars appeared first in \cite{Lewis:2011zb,Hietanen:2014xca} and further developed in \cite{Arthur:2016dir,Bennett:2017kga,Lee:2017uvl,Drach:2017btk}. The analytic ultraviolet and perturbative conformal structure and fate of these type of theories has been carefully analyzed in \cite{Hansen:2017pwe,Einhorn:2017jbs}.

At the same time there has been a growing interest in the study of flavor physics as a means to provide insight into new physics.
Given the lack of direct evidence for new particles at LHC so far, flavor physics provides a unique opportunity to probe energy
scales not accessible directly.
Flavor observables are also well known to impose stringent constraints on models with new composite dynamics
\cite{Csaki:2008zd,Agashe:2008uz,Barbieri:2012tu}.
Interestingly, several deviations from SM expectations have been observed in flavor physics in recent years. Most notably,
hints for a violation of lepton flavor universality in $b\to s\ell^+\ell^-$ transitions with $\ell=e$ vs. $\mu$
\cite{Aaij:2014ora,Aaij:2017vbb},
and independent hints for a violation of lepton flavor universality in $b\to c\ell\nu$ transitions with $\ell=\tau$ vs.\ $\mu$ or $e$
\cite{Lees:2013uzd,Huschle:2015rga,Aaij:2015yra,Sato:2016svk,Hirose:2016wfn,Aaij:2017uff}.
If confirmed, these deviations would constitute unambiguous evidence of physics beyond the SM.
It is thus important to look for models that can accommodate these anomalies.

The aim of this paper is to perform a comprehensive study of flavor constraints
on MFPC
and to investigate whether it can explain the aforementioned ``flavor anomalies''.
The remainder of the paper is organized as follows.
In section~\ref{sec:model}, we review the MFPC model and fix our notation.
In section~\ref{sec:weh}, we discuss all relevant low-energy precision constraints in our analysis and present approximate analytical
formulae for the MFPC contributions.
Section~\ref{sec:numerics} contains the description of our strategy and the discussion of the results of our global numerical
analysis of flavor in MFPC.
Section~\ref{sec:conclusions} contains our conclusions.

\section{Minimal Fundamental Partial Compositeness}\label{sec:model}
\begin{table}[b!]
	\centering
	\begin{tabular}{|l|c c c | c c|}
		\hline
		& $ \Fupdn $ & $ \Fupbar $ & $ \Fdnbar $ & $ \tcs_q $ & $ \tcs_l $\\ \hline \rule{0pt}{.1em}
		$ G_\mathrm{SM} $ & $ \left(1,2,0\right) $ & $ \left(1,1, -\tfrac{1}{2} \right) $ & $ \left(1,1, \tfrac{1}{2}\right) $ & $ 3 \times \left(\overline{3}, 1, -\tfrac{1}{6}\right) $ & $ 3\times \left(1, 1, \tfrac{1}{2}\right) $ \\
		TC symmetries & \multicolumn{3}{c|}{$ 4_\tcf \otimes N_\TC $} & \multicolumn{2}{c|}{$ 24_\tcs \otimes N_\TC $ } \\ \hline
	\end{tabular}
	\caption{The table summarizes the new BSM states with their representation under $ G_\SM $. Furthermore, it provides their flavor symmetry in the absence of SM interactions and their representation of $ G_\TC $. The left part lists the technifermions, $ \tcf $, and the right part the techniscalars, $ \tcs $.}
	\label{tab:TC_states}
	\vspace{.5cm}
	\begin{tabular}{|l|c c c| c c c|}
		\hline
		SM fermion & $ Q $ & $ \ubar $ & $ \dbar $ & $ L $ & $ \nubar $ & $ \ebar $ \\ \hline
		\rule{0pt}{13pt} $ G_\mathrm{SM} $ & $ \left(3, 2 , \tfrac{1}{6}\right) $ & $ \left(\overline{3}, 1, -\tfrac{2}{3}\right) $ & $ \left(\overline{3}, 1, \tfrac{1}{3}\right) $ & $ \left(1,2,- \tfrac{1}{2} \right) $ & $ \left(1,1,0\right) $ & $ \left(1, 1, 1\right) $ \\[.2em] \hline
	\end{tabular}
	\caption{A summary of the two-component SM spinors and their quantum numbers.}
	\label{tab:SM_fermions}
\end{table}

\noindent
In the MFPC model, the SM is appended with a new fundamental sector featuring a strong TC force.
This sector contains both TC fermions, or technifermions, $ \tcf $, and TC scalars, or techniscalars, $ \tcs $, which are charged under TC and will form bound states below the TC confining scale.
In particular, the Higgs boson will be realized as a bound state of technifermions, while the partial compositeness mechanism is realized through a mixing between the SM fermions and fermionic bound states consisting of both technifermions and techniscalars.
The full kinetic term of the new TC sector with both technifermions and techniscalars transforming in the fundamental, pseudoreal representation of $ G_\TC = \Sp(N_\TC) $ is then given by
	\begin{equation}
	\mathcal{L}_{\mathrm{kin}} = -\tfrac{1}{4} \mathcal{G}^{a}_{\mu\nu} \mathcal{G}^{a\mu\nu} + i \tcf^\dagger \bar{\sigma}^\mu D_\mu \tcf - \left(\tfrac{1}{2} \tcf\transpose m_\tcf \epsilon_\TC \tcf \hc\right) + \left(D_\mu \tcs\right)^{\dagger} \left(D^{\mu} \tcs\right) -  \tcs^{\dagger} m^2_\tcs \tcs.
	\label{eq:L_kin}
	\end{equation}
For there to be mixing between the fermionic bound states and SM fermions, the model includes fundamental Yukawa interactions between the TC and the SM sectors.
This requires the TC particles to carry SM charges and the technifermions are taken to be in a vector-like representation of $ G_\SM $ to avoid gauge anomalies. 
Given these constraints, the minimal content of new TC matter is given in Table \ref{tab:TC_states}. The most general fundamental Yukawa interaction between the new TC sector and the elementary (SM) fermions are then given by
	\begin{equation}
	\begin{split}
	\mathcal{L}_{\mathrm{yuk}} = \, &
	y_Q\, Q_{\alpha}\, \tcs_q\epsilon_{\TC} \Fupdn^{\alpha}
	-y_\ubar\, \ubar\, \tcs_q^{\ast} \Fdnbar
	+ y_\dbar\, \dbar\, \tcs_q^{\ast} \Fupbar\\
	& + y_L\, L_\alpha\, \tcs_l \epsilon_{\TC} \Fupdn^\alpha
	- y_\nubar\, \nubar\, \tcs_l^{\ast} \Fdnbar
	+ y_\ebar\, \ebar\, \tcs_l^{\ast} \Fupbar
	-\tilde{y}_\nubar\, \nubar\, \tcs_l \Fupbar
	\hc
	\end{split}
	\label{eq:Lfund_y}
	\end{equation}
where $ \alpha $ is an $ \SU(2)_\LL $ index (here implicitly contracted using the $ \SU(2) $-invariant tensor).
For completeness, we list the quantum numbers of SM fermions in Table \ref{tab:SM_fermions}, too. Giving masses to all three generation of SM fermions, i.e.\ avoiding vanishing eigenvalues in the mass matrices, requires three generations of techniscalars, such that the total TC particle content is $ 12 N_\TC $ techniscalars and $ 4 N_\TC $ technifermions.
In this construction, the fundamental Yukawa couplings $ y_f $ are to be understood as $ 3 \times 3 $ matrices.
On a final note, the right-handed neutrinos are assumed to be irrelevant for the low-energy flavor observables we consider in our analysis.
We consequently neglect their effects by taking $ y_\nubar = \tilde{y}_\nubar =0 $ in the following.

\subsection{Global flavor symmetries and electroweak symmetry breaking}
As discussed in~\cite{Cacciapaglia:2017cdi}, in the absence of the mass terms $ m_{\tcf,\tcs} $, the technifermions satisfy an $ \SU(4)_\tcf $ symmetry, while the techniscalars have an enhanced $ \Sp(24)_\tcs $ global flavor symmetry.
More generally, the global symmetries of eq.~\eqref{eq:L_kin} are explicitly broken both by SM interactions and by the mass terms. It is, however, assumed that the strong dynamics will dominate the new physics at the new composite scale $ \Lambda_\TC $, while SM interactions remain subdominant. The symmetries of the strong sector are thus expected to be approximately preserved in the low-energy effective theory. Therefore, the TC particles are conveniently arranged as
	\begin{equation}
	\tcf^{a} \in 4_\tcf \otimes N_\TC \andeq \Phi^{i} = \begin{pmatrix} \tcs \\ -\epsilon_\TC \tcs^{\ast} \end{pmatrix} \in 24_\tcs \otimes N_\TC,
	\end{equation}
where $ a $ is an $ \SU(4)_\tcf $ and $ i $ is an $ \Sp(24)_\tcs $  index. In terms of $\tcf^{a}$ and $\Phi^{i}$, the fundamental Yukawa interactions of eq.~\eqref{eq:Lfund_y} are given by
	\begin{equation}
	\mathcal{L}_{\mathrm{yuk}} = - \spur{i}{a} \epsilon_{ij} \Phi^{j} \epsilon_\TC \tcf^{a} \hc,
	\label{eq:Lyuk_sym}
	\end{equation}
where the spurion field $ \psi $ consists of SM fermions and Yukawa matrices:
 	\begin{equation} \label{eq:spurionpsi}
 	\spur{i}{a} \equiv \left(\Psi \, y\right)^{i} \phantom{}_{a} \in \overline{4}_{\tcf} \otimes 24_{\tcs}.
 	\end{equation}
As always, the benefit of the spurionic fields are that they may be included systematically in the low energy EFT to control the degree of breaking of the approximate flavor symmetries. In particular, the spurionic fields carry chiral dimension from the perspective of systematic power counting, so operators with more insertions are suppressed. Note that the SM fermions only couple directly to the strong sector through $ y_f $, and so they will always appear in the combination $ \psi $.
For the purpose of this analysis, we will work in the limit of a flavor-trivial scalar mass matrix (proportional to unity). More generally, a small but non-vanishing mass matrix, $ m_\tcs^2 \ll \Lambda_\TC^2 $, can be included systematically in the low-energy effective theory, but would not contribute to the order considered in this work.

The symmetry breaking of the model begins at the composite scale, $ \Lambda_\TC $, of the TC dynamics. At this scale, the fermions are expected to form a condensate
	\begin{equation}
	\brakets{\tcf^{a} \epsilon_\TC \tcf^{b} } = \Lambda_\TC f_\TC^2 \Sigma_{\theta}^{a b},
	\end{equation}
thereby spontaneously breaking the global $ \SU(4)_\tcf $ symmetry to an $ \Sp(4) $ subgroup. Here $ \Sigma_\theta $ is an antisymmetric matrix determining the alignment of the $ \Sp(4) $ stability group in $ \SU(4) $, and $ f_\TC \sim \Lambda_\TC/4\pi $ is the decay constant of the Nambu-Goldstone Bosons (NGBs) associated to the spontaneous breaking.
In the case of an exact global $ \SU(4)_\tcf $ symmetry, making distinctions between different alignments is pointless (and futile).
However, in the realistic case, the EW gauge group is embedded into the $ \SU(4)_\tcf $ group thus introducing a preferred direction for the vacuum alignment. The physical vacuum alignment is then parametrized using an angle $ \theta $ such that
	\begin{equation}
	\Sigma_\theta = c_\theta \begin{pmatrix}
	i \sigma_2 & 0 \\ 0 & - i \sigma_2
	\end{pmatrix} + s_\theta \begin{pmatrix}
	0 & \mathds{1}_2 \\ - \mathds{1}_2 & 0
	\end{pmatrix},
	\end{equation}
where $ c_\theta = \cos \theta $ and $ s_\theta = \sin \theta $~\cite{Galloway:2010bp}. Here $ c_\theta = 1 $ corresponds to a vacuum which preserves the EW gauge symmetry, whereas $ s_\theta = 1 $ leaves it maximally broken.

The NGBs of the $\SU(4)_\mathcal{F}\to \Sp(4)$ symmetry breaking are parametrized by fluctuations around the vacuum $\Sigma_\theta$ in terms of the matrix
	\begin{equation}
	\Sigma(x) = \exp\left[i \frac{2\sqrt{2}}{f_\TC} \Pi_i(x) X^i_\theta \right] \Sigma_\theta.
	\end{equation}
Here $X^i_\theta$ are the broken generators%
\footnote{For the NGBs to parametrize the fluctuations around the actual $\theta$-dependent vacuum $\Sigma_\theta$, the parametrization of the broken generators also depends on $\theta$ (cf.~\cite{Cacciapaglia:2014uja}).} 
of $\SU(4)_\mathcal{F}$, $ \Pi_{1,2,3} $ are identified with the EW NGBs, $ \Pi_4 $ with the Higgs boson, and $ \Pi_5 $ is an SM singlet. 
As we will describe in more detail in the next section, physics at low energies can be described using an EFT. In this effective description, the NGBs appear through the leading-order (LO) kinetic term
	\begin{equation}
	\mathcal{L}_\mathrm{EFT} \supset \dfrac{f_\TC^2}{8} \tr\left[ D_\mu \Sigma^\dagger \, D^\mu \Sigma \right],
	\label{eq:L_kin_GBs}
	\end{equation}
which also gives rise to mass terms for the EW gauge bosons. In particular, recovering the experimental masses yields the relation $ v_\EW = s_\theta f_\TC $.

A radiatively generated potential promotes the NGBs to pNGBs and determines the actual alignment of the vacuum.
These radiative effects are due to terms in the fundamental Lagrangian that explicitly break the global symmetry: fundamental fermion masses, EW gauge couplings, and Yukawa couplings.
Identifying the Higgs with the $ \Pi_4 $ pNGB only makes sense in the case $ 0 < s_\theta \ll 1 $ (cf.~\cite{Galloway:2010bp,Cacciapaglia:2014uja}).
For the model considered here, contributions to the effective potential are discussed in~\cite{Cacciapaglia:2017cdi} and a small value for $s_\theta$ can be realized.
We therefore assume in the following that $ 0 < s_\theta \ll 1 $ and allow for different values of $s_\theta$ in our numerics by varying $f_\TC$ while keeping $v_\EW$ fixed (cf.\ section~\ref{sec:numerics_parameters}). 
Of the pNGB fields, only $ \Pi_5 $ is new as compared to the SM. It generically has a mass $ m = m_h / s_\theta $ and does not have a Yukawa coupling to the SM fermions at leading order \cite{Galloway:2010bp}. For this reason we will ignore it in our analysis. 

\subsection{Effective theory at the electroweak scale} \label{sec:effective_model}
The TC condensation scale $ \Lambda_\TC $ is expected to be large compared to the EW scale, such that there is a clear hierarchy $ \mathrm{v_\EW} \ll \Lambda_\TC $. The effects of the new composite dynamics on SM physics at the EW scale can thus be described by an EFT in a controlled manner, where the effective degrees of freedom include the SM fermions and gauge bosons, and the pNGBs $ \Pi_i $. Meanwhile, the effects of physics above $ \Lambda_\TC $ are included in effective operators consistent with the symmetries of the underlying dynamics.
The resulting theory, which we will refer to as the MFPC-EFT, was determined in detail in Ref.~\cite{Cacciapaglia:2017cdi}, and here we just present the operators of relevance for our analysis. 
The effective Lagrangian can be written as
	\begin{equation}
	\mathcal{L}_{\mathrm{EFT}} = \mathcal{L}_{\SM-\mathrm{Higgs}} + \sum_{A} C_A \, \mathcal{O}_A + \left(\sum_{A}C'_A\, \mathcal{O}'_A \hc \right),
	\end{equation}
where the new physics is contained in the $ \mathcal{O}^{(\prime)}_A $ operators. The normalization of the effective operators is due to symmetry factors and power counting for strongly interacting electroweak EFTs~\cite{Buchalla:2013eza}\footnote{In contrast to Ref.~\cite{Cacciapaglia:2017cdi} we have not rescaled the fundamental Yukawas $ y_f $.}.
The strong coefficients $ C^{(\prime)}_A $ are determined by the underlying TC dynamics, and expected to be $ \mathcal{O}(1) $ with the present choice of operator normalization.

The leading-order operator with just two SM fermions in the effective theory is given by
	\begin{equation} \label{eq:OYuk}
	\mathcal{O}_{\mathrm{Yuk}} = -\dfrac{f_\TC}{8 \pi} \, (\spur{i_1}{a_1} \spur{i_2}{a_2})\, \Sigma^{a_1 a_2} \epsilon_{i_1 i_2 }.
	\end{equation}
It is responsible for giving masses to the SM fermions and also provides a coupling to the Higgs boson (hence its name). In the flavor analysis of the model, this operator constrains the fundamental Yukawas $ y_f $ to reproduce the SM masses and the Cabibbo–Kobayashi–Maskawa (CKM) matrix.

Of particular relevance for the purpose of flavor physics are four-fermion operators induced by the underlying dynamics. They are completely described by the set of self-hermitian operators
	\begin{align}
	\mathcal{O}^{1}_{4f} &= \dfrac{1}{64\pi^2 \Lambda_\TC^2} (\spur{i_1}{a_1} \spur{i_2}{a_2} ) (\spurbar{i_3}{a_3} \spurbar{i_4}{a_4} ) \Sigma^{a_1 a_2} \Sigma^\dagger_{a_3 a_4} \epsilon_{i_1 i_2} \epsilon_{i_3 i_4}\ ,\label{eq:fourfermion1} \\
	\mathcal{O}^{2}_{4f} &= \dfrac{1}{64 \pi^2 \Lambda_\TC^2} (\spur{i_1}{a_1} \spur{i_2}{a_2} ) (\spurbar{i_3}{a_3} \spurbar{i_4}{a_4} ) \left(\delta^{a_1}_{\enspace a_3} \delta^{a_2}_{\enspace a_4} - \delta^{a_1}_{\enspace a_4} \delta^{a_2}_{\enspace a_3} \right) \epsilon_{i_1 i_2} \epsilon_{i_3 i_4}\ , \\
	\mathcal{O}^{3}_{4f} &= \dfrac{1}{64\pi^2 \Lambda_\TC^2} (\spur{i_1}{a_1} \spur{i_2}{a_2} ) (\spurbar{i_3}{a_3} \spurbar{i_4}{a_4} ) \Sigma^{a_1 a_2} \Sigma^\dagger_{a_3 a_4} \left(\epsilon_{ i_1 i_4} \epsilon_{ i_2 i_3} - \epsilon_{ i_1 i_3} \epsilon_{ i_2 i_4} \right)\ ,	\\
	\mathcal{O}^{4}_{4f} &= \dfrac{1}{64 \pi^2 \Lambda_\TC^2} (\spur{i_1}{a_1} \spur{i_2}{a_2} ) (\spurbar{i_3}{a_3} \spurbar{i_4}{a_4} ) \left( \delta^{a_1}_{\enspace a_3} \delta^{a_2}_{\enspace a_4} \epsilon_{ i_1 i_3} \epsilon_{ i_2 i_4} + \delta^{a_1}_{\enspace a_4} \delta^{a_2}_{\enspace a_3} \epsilon_{ i_1 i_4} \epsilon_{ i_2 i_3}\right)\ , \\
	\mathcal{O}^{5}_{4f} &= \dfrac{1}{64 \pi^2 \Lambda_\TC^2} (\spur{i_1}{a_1} \spur{i_2}{a_2} ) (\spurbar{i_3}{a_3} \spurbar{i_4}{a_4} ) \left( \delta^{a_1}_{\enspace a_3} \delta^{a_2}_{\enspace a_4} \epsilon_{ i_1 i_4} \epsilon_{ i_2 i_3} + \delta^{a_1}_{\enspace a_4} \delta^{a_2}_{\enspace a_3} \epsilon_{ i_1 i_3} \epsilon_{ i_2 i_4}\right)\ , \label{eq:fourfermion5}
	\end{align}
and the complex operators
	\begin{align}
	\mathcal{O}^{6}_{4f} &= \dfrac{1}{128 \pi^2 \Lambda_\TC^2} (\spur{i_1}{a_1} \spur{i_2}{a_2} ) (\spur{i_3}{a_3} \spur{i_4}{a_4} )  \Sigma^{a_1 a_2} \Sigma^{a_3 a_4} \epsilon_{i_1 i_2} \epsilon_{i_3 i_4}\,, \label{eq:fourfermion6} \\
	\mathcal{O}^{7}_{4f} &= \dfrac{1}{128 \pi^2 \Lambda_\TC^2} (\spur{i_1}{a_1} \spur{i_2}{a_2} ) (\spur{i_3}{a_3} \spur{i_4}{a_4} )  \left(\Sigma^{a_1 a_4} \Sigma^{a_2 a_3} - \Sigma^{a_1 a_3} \Sigma^{a_2 a_4}\right) \epsilon_{i_1 i_2} \epsilon_{i_3 i_4}\,,   \\
	\mathcal{O}^{8}_{4f} &= \dfrac{1}{128 \pi^2 \Lambda_\TC^2} (\spur{i_1}{a_1} \spur{i_2}{a_2} ) (\spur{i_3}{a_3} \spur{i_4}{a_4} )  \Sigma^{a_1 a_2} \Sigma^{a_3 a_4} \left(\epsilon_{i_1 i_4} \epsilon_{i_2 i_3} - \epsilon_{i_1 i_3} \epsilon_{i_2 i_4}\right)\, . \label{eq:fourfermion8}
	\end{align}

The TC sector is also responsible for modifying the couplings between SM fermions and SM gauge bosons. It induces the operator
	\begin{equation} \label{eq:OPif}
	\mathcal{O}_{\Pi f} = \dfrac{i}{32\pi^2 }(\spurbar{i_1}{a_1} \bar{\sigma}_\mu \spur{i_2}{a_2} )\  \Sigma _{a_1 a_3}^\dag  \overleftrightarrow{D}^\mu \Sigma ^{a_3 a_2}\  \epsilon _{i_1 i_2}\,,
	\end{equation}
that modifies the couplings of the weak gauge bosons and is mainly constrained by LEP measurements of the $Z$ branching ratios (cf.\ section~\ref{sec:constraints_Z}).

\section{Low-energy signals from the Weak Effective Hamiltonian}\label{sec:weh}
\begin{figure}[t]
	\includegraphics[width=.8\textwidth]{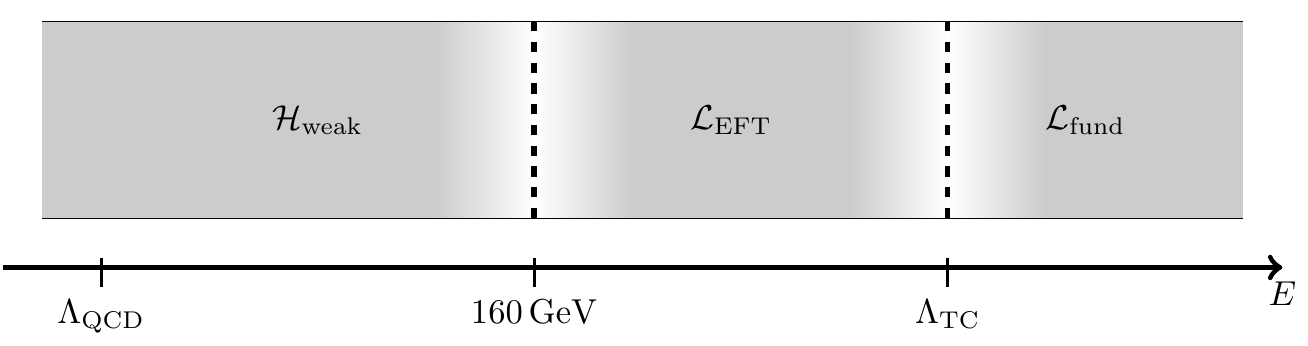}
	\caption{Schematic representation of the theory descriptions employed in our analysis. The fundamental theory in the UV is in principle matched to the MFPC-EFT at the scale of compositeness, $ \Lambda_\TC $, although without Lattice results we only posses naive estimates for the coefficients. Flavor physics is most conveniently described by the Weak effective Hamiltonian at low energies. We match the MFPC-EFT with the WEH at the scale $ \SI{160}{GeV} $.}
	\label{fig:EFTs}
\end{figure}

To determine the effect of the MFPC model on low-energy observables, we follow the usual approach and derive its consequences on the Weak Effective Hamiltonian (WEH), $ \mathcal{H}_\mathrm{weak} $.
As illustrated in Fig.~\ref{fig:EFTs}, we describe the physics at intermediate scales between $\Lambda_\TC$ and the low energy regime using the effective theory $ \mathcal{L}_\mathrm{EFT} $ as discussed above.
At the the scale of $ \SI{160}{GeV} $, $W$, $Z$, $t$ and the pNGBs $\Pi_i$ are integrated out and $ \mathcal{L}_\mathrm{EFT} $ is matched to the WEH $ \mathcal{H}_\mathrm{weak} $.
The benefit of this procedure is a controlled treatment of the (approximate) UV symmetries, which we can now trace to correlated operators in $ \mathcal{H}_\mathrm{weak} $.

\subsection{Matching the MFPC-EFT to the Weak Effective Hamiltonian} \label{sec:sm_yukawas}
Among the MFPC-EFT operators,
only $ \mathcal{O}_\mathrm{Yuk} $ contains terms that are also present in the SM.
These are the fermion-Higgs couplings and the fermion mass terms.
In unitary gauge, we have
	\begin{equation}
	C_\mathrm{Yuk} \mathcal{O}_\mathrm{Yuk} =-\sum_{f\in\{u,d,e\}} \dfrac{C_\mathrm{Yuk}\, s_\theta\, f_\TC}{4\pi} \, (y_{f}\transpose\, y_{\bar{f}})_{ij} \left(f_{i} \bar{f}_{j} \right) \left(1 + \dfrac{c_\theta h} {v_{\mathrm{EW}}} + \dots \right)
	\label{eq:SM_masses}
	\end{equation}
ignoring nonlinear terms in the pNGBs.
We employ a compact notation where the fundamental Yukawa couplings of the $\SU(2)_\LL$ doublets are labeled by the names of their doublet components, i.e.\ we use $y_{Q}=y_{u}=y_{d}$ and $y_{L}=y_{e}=y_{\nu}$.
From the mass term, one may identify the mass matrices of the SM fermions
	\begin{equation}
	m_{f,ij} = \dfrac{C_\mathrm{Yuk}\, s_\theta\, f_\TC}{4\pi}\,  \left( y_{f}\transpose\, y_{\bar{f}} \right)_{ij}.
	\end{equation}
The WEH is defined in the mass basis where $m_{f,ij}$ has been diagonalized by a biunitary transformation
\begin{equation}\label{eq:mass_basis}
 m_{f}^{\text{diag}}=
 U_{f}^{\transpose}
 \,
 m_{f}
 \,
 U_{\bar{f}},
 \quad\quad
 f\in\{u,d,e\},
\end{equation}
which defines the unitary matrices $U_{f}$ and $U_{\bar{f}}$.
These matrices appear in the Wilson coefficients of the WEH in the following combinations\footnote{Since we treat neutrinos as massless, the charged lepton mass matrix can be chosen to be diagonal already in the gauge-basis, such that $U_{e}=U_{\nu}=U_{\bar{e}}=\mathds{1}_3$ and the contribution of $U_{e}$, $U_{\nu}$ and $U_{\bar{e}}$ to the Wilson coefficients is trivial.}:
\begin{itemize}
 \item In the CKM matrix defined by
 \begin{equation}\label{eq:CKM}
  V = U_{u}^\dagger\,U_{d}.
 \end{equation}
 \item In a product of two fundamental Yukawa matrices where one of them is complex conjugated and the other is not:
 \begin{equation}
  X_{f_1 f_2} = \dfrac{1}{4\pi} \, U_{f_1}^\dagger\,y_{f_1}^\dagger\,y_{f_2}\,U_{f_2},
  \quad\quad
  X_{f_1 f_2}^* = X_{f_2 f_1}\transpose.
 \end{equation}
 \item In a product of two fundamental Yukawa matrices where both of them are either unconjugated or conjugated:
 \begin{equation}
  Y_{f_1 f_2} =\dfrac{1}{4\pi} \, U_{f_1}\transpose\,y_{f_1}\transpose\,y_{f_2}\,U_{f_2},
  \quad\quad
  Y_{f_1 f_2}^* = \dfrac{1}{4\pi} \, U_{f_1}^\dagger\,y_{f_1}^\dagger\,y_{f_2}^*\,U_{f_2}^*.
 \end{equation}
\end{itemize}
For the last two cases, $f_1$ and $f_2$ denote a SM fermion, i.e. $f_1,f_2 \in \{u,d,e,\nu,\bar{u},\bar{d},\bar{e}\}$.
Using the definition of $Y_{f_1 f_2}$, the fermion mass matrices in the mass basis can be written as
\begin{equation}
 m_{f}^{\text{diag}}=C_\mathrm{Yuk}\, s_\theta\, f_\TC\, Y_{f \bar{f}}
\end{equation}
and the mass basis SM Yukawa couplings $Y_f^{\rm SM}$ can be identified as
\begin{equation}
 Y_{f}^{\rm SM} = \sqrt{2}\,C_\mathrm{Yuk}\,Y_{f \bar{f}}.
\end{equation}
Apart from $\mathcal{O}_\mathrm{Yuk}$, all operators of $ \mathcal{L}_\mathrm{EFT}$ describe pure NP effects not present in the SM.
As such, they lead to deviations of the WEH Wilson coefficients with respect to the SM contributions.

The four-fermion operators $\mathcal O^i_{4f}$ can be readily matched to the WEH by summing over the global $ \SU(4)_\tcf $ and $ \Sp(24)_\tcs $ indices.
For this purpose, we note that the spurion field $ \psi $ assumes the value
	\begin{equation}
	\spur{i}{a}=\begin{pmatrix}
	0 & 0 & y_{\bar{d}}\, \bar{d} & -y_{\bar{u}}\, \bar{u} \\
	0 & 0 & y_{\bar{e}}\, \bar{e} &
	0
	\\
	y_Q\, d & - y_Q\, u & 0 & 0 \\
	y_L\, e & - y_L\, \nu &
	0
	& 0
	\end{pmatrix},
	\end{equation}
keeping the $ \SU(3)_c $ and SM generation part of the $ \Sp(24)_\tcs $ index implicit.
The spinors as well as the fundamental Yukawa couplings are rotated to the mass bases via the unitary matrices defined in eq. (\ref{eq:mass_basis}).
The resulting four-fermion operators are still expressed in the two-component chiral Weyl spinor notation employed in sec \ref{sec:effective_model}.
We thus subsequently apply an assortment of Fierz identities to match them to the WEH basis defined in terms of 4-component Dirac spinors.

Besides the four-fermion operators in $ \mathcal{L}_\mathrm{EFT}$, an important role in our analysis is played by the operator $\mathcal{O}_{\Pi f}$.
Modifying the couplings of weak gauge bosons to SM fermions, it yields NP contributions to four-fermion operators in the WEH when integrating out the $W$ and $Z$ bosons.
For the matching, we first derive the $W$- and $Z$-couplings contained in $\mathcal{O}_{\Pi f}$.
We then integrate out the $W$ and $Z$ bosons, yielding new four-fermion operators below the EW scale from tree-level weak gauge boson exchange, where either one or both ends of the gauge boson propagator couples to the SM fermions via the NP coupling induced by $\mathcal{O}_{\Pi f}$.
These four-fermion operators are then matched to the WEH by applying the same steps as for the four-fermion operators $\mathcal O^i_{4f}$ described above.

Since the operator $\mathcal{O}_\mathrm{Yuk}$ will slightly modify the Higgs couplings to SM fermions, it leads to NP contributions to four-fermion operators in the WEH when integrating out the Higgs.
However, these operators are always flavor-diagonal and subleading in an expansion in $ s_\theta^2 $ and we will therefore neglect their contributions.

\subsection{Constraints from EW scale physics}
In addition to contributing to four-fermion operators in the WEH, the operators $\mathcal{O}_\mathrm{Yuk}$ and $\mathcal{O}_{\Pi f}$ also affect observables at the EW scale.
The modified Higgs couplings contained in the former are constrained by measurements at the LHC and the new couplings of weak gauge bosons to SM fermions induced by the latter are constrained by $Z$-boson observables measured at LEP.

\subsubsection{Higgs boson couplings}
A pNGB Higgs boson in the $ \SU(4)/\Sp(4) $ breaking pattern has non-standard couplings to the SM particles as compared to the SM Higgs~\cite{Galloway:2010bp,Cacciapaglia:2014uja}. The modification of the Higgs coupling to fermions can be read directly off eq.~\eqref{eq:SM_masses}, and the single couplings to the weak gauge bosons may be found by expanding the kinetic term of eq.~\eqref{eq:L_kin_GBs}. One finds
	\begin{equation} \label{eq:Higgs_couplings}
	g_{ffh} = c_\theta g_{ffh}^{\mathrm{SM}}\,, \quad g_{ZZh} = c_\theta g_{ZZh}^{\mathrm{SM}}\,, \quad g_{WWh} = c_\theta g_{WWh}^{\mathrm{SM}}\,.
	\end{equation}
The resulting collider constraints have already been discussed in depth in the existing literature, see e.g.\ \cite{Sanz:2017tco}, so we will merely note that the strongest individual constraint comes from the Higgs coupling to the $ Z $ boson. The combined ATLAS and CMS analysis \cite{Khachatryan:2016vau}, using the Run~I LHC data, yields the bound 	\begin{equation}
	s_\theta < 0.44 \quad @ \, 68\% \, \mathrm{CL}
	\end{equation}
just from the $ hZZ $ coupling. In our analysis, we will only consider points with $ f_\TC \geq \SI{1}{TeV} $ ($ s_\theta < 0.25 $), and the constraints coming from Higgs physics will be satisfied in all cases.

\subsubsection{\texorpdfstring{$Z$}{Z} boson couplings}\label{sec:constraints_Z}
The NP couplings of the $Z$ boson to SM fermions that are induced by $\mathcal{O}_{\Pi f}$ can be expressed as
\begin{equation}
C_{\Pi f}\,\mathcal{O}_{\Pi f} \supset
\sum_{f\in\{u,d,e,\nu\}}
 \frac{g}{c_W} Z_\mu \left( {\delta g}_{f_L}^{ij}\, \bar{f}_{L}^i\, \gamma^\mu\, f_{L}^j
 + {\delta g}_{f_R}^{ij}\, \bar{f}_{R}^i\, \gamma^\mu\, f_{R}^j\, \right),
\end{equation}
where the deviations  ${\delta g}_{f_L}^{ij}$ and ${\delta g}_{f_R}^{ij}$ from the SM $Z$ couplings are given by
\begin{equation}\label{eq:Z_couplings}
 \begin{aligned}
{\delta g}_{u_L}^{ij} &=
 +\frac{C_{\Pi f}}{8\pi}\,s_\theta^2\, \big(X_{uu}\big)_{ij}\,,
 \quad\quad
{\delta g}_{u_R}^{ij} =
 -\frac{C_{\Pi f}}{8\pi}\,s_\theta^2\, \big(X_{\bar{u}\bar{u}}^*\big)_{ij}\,,
 \\
{\delta g}_{d_L}^{ij} &=
 -\frac{C_{\Pi f}}{8\pi}\,s_\theta^2\, \big(X_{dd}\big)_{ij}\,,
 \quad\quad
{\delta g}_{d_R}^{ij} =
 +\frac{C_{\Pi f}}{8\pi}\,s_\theta^2\, \big(X_{\bar{d}\bar{d}}^*\big)_{ij}\,,
 \\
{\delta g}_{e_L}^{ij} &=
 -\frac{C_{\Pi f}}{8\pi}\,s_\theta^2\, \big(X_{ee}\big)_{ij}\,,
 \quad\quad
{\delta g}_{e_R}^{ij} =
 +\frac{C_{\Pi f}}{8\pi}\,s_\theta^2\, \big(X_{\bar{e}\bar{e}}^*\big)_{ij}\,,
 \\
{\delta g}_{\nu_L}^{ij} &=
 +\frac{C_{\Pi f}}{8\pi}\,s_\theta^2\, \big(X_{\nu\nu}\big)_{ij}\,,
 \quad\quad
{\delta g}_{\nu_R}^{ij} = 0\,.
 \end{aligned}
\end{equation}
The flavor-diagonal terms modify the $Z$ partial widths measured at LEP.
To reproduce the correct top quark mass, the fundamental Yukawa couplings of the third generation quark doublet are usually large\footnote{To some degree, large fundamental Yukawa couplings of the top quark singlet can ease the requirement of large doublet couplings. However, even for singlet couplings of $\mathcal{O}(4\pi)$, the doublet couplings have to be $\mathcal{O}(1)$ and are thus never small.}. This can yield a sizable contribution to the $Z b_L b_L$ coupling and thus be in conflict with LEP data.
In effective models of partial compositeness that satisfy all EW precision constraints, this problem is usually avoided by a custodial protection of the $Z b_L b_L$ coupling~\cite{Agashe:2006at,Contino:2006qr}.
Since the MFPC model does not feature a protection of this kind\footnote{Possible FPC models that include a custodial protection of the $Z b_L b_L$ coupling are discussed in~\cite{Sannino:2016sfx}.}, the LEP measurements of partial widths of the $Z$ boson are important constraints that have to be taken into account.
To this end, we calculate the following observables for each parameter point,
\begin{equation}\label{eq:Z_observables_quarks}
 R_b=\frac{\Gamma(Z\to b\bar{b})}{\Gamma(Z\to q\bar{q})},
 \quad\quad
 R_c=\frac{\Gamma(Z\to c\bar{c})}{\Gamma(Z\to q\bar{q})},
\end{equation}
\begin{equation}\label{eq:Z_observables_leptons}
 R_e=\frac{\Gamma(Z\to q\bar{q})}{\Gamma(Z\to e\bar{e})},
 \quad\quad
 R_\mu=\frac{\Gamma(Z\to q\bar{q})}{\Gamma(Z\to \mu\bar{\mu})},
 \quad\quad
 R_\tau=\frac{\Gamma(Z\to q\bar{q})}{\Gamma(Z\to \tau\bar{\tau})},
\end{equation}
where $\Gamma(Z\to q\bar{q})$ implies a sum over all quarks except the top.
We include higher order electroweak corrections~\cite{Freitas:2014hra}
as well as the leading order QCD correction~\cite{Chetyrkin:1994js} to reproduce the correct SM predictions in the limit $C_{\Pi f}=0$.

\begin{table}[tbp]
\centering
\begin{tabular}{lll}
\hline
Observable & measurement \\
\hline
$R_e$	& 20.804(50) & \cite{ALEPH:2005ab}  \\
$R_\mu$	& 20.785(33) & \cite{ALEPH:2005ab}  \\
$R_\tau$& 20.764(45) & \cite{ALEPH:2005ab}  \\
$R_b$	& 0.21629(66)& \cite{ALEPH:2005ab}  \\
$R_c$	& 0.1721(30) & \cite{ALEPH:2005ab}  \\
\hline
\end{tabular}
\caption{Experimental values of $Z$ boson partial width ratios used in our numerical analysis.}
\label{tab:experimental_values}
\end{table}

\subsubsection{Electroweak precision tests}
In addition to the above described observables, the model is constrained by EW precision data in form of the $ S $ and $ T $ parameters~\cite{Peskin:1991sw}. There are contributions to the $S$, $T$ parameters due to non-standard couplings between the SM particles, which will result in contributions to the EW vacuum polarizations different from the SM prediction. At leading order in the MFPC-EFT, only the Higgs coupling is different from the SM, cf.\ eq.\ \eqref{eq:Higgs_couplings}, and so only the pNGB loops will give loop contributions to $S$, $T$. It was shown in Ref.~\cite{Foadi:2012ga} that this results in contributions\footnote{$ f(x) = \frac{2x^2+x^4-3x^6 +(9x^4 + x^6)\log x}{(1-x^2)^3} $ is a loop function. }
	\begin{align}
	S_\mathrm{IR} &= S^\mathrm{MFPC}_\mathrm{pNGB} - S^\mathrm{SM}_\mathrm{Higgs} = \dfrac{s_\theta^2}{12\pi} \left[f(m_z/m_h) + \log\dfrac{\Lambda_\TC^2}{m_h^2} + \dfrac{5}{6},  \right], \\
	T_\mathrm{IR} &= T^\mathrm{MFPC}_\mathrm{pNGB} - T^\mathrm{SM}_\mathrm{Higgs} = - \dfrac{3 s^2_\theta}{16 \pi c_w^2} \log \dfrac{\Lambda^2_\TC }{m_h^2 }.
	\end{align}
Here the divergences have been replaced with $ \Lambda_\TC $, as they will be absorbed into counter terms at next-to-leading order (NLO). Additionally, the $S$, $T$ parameters will receive contributions from physics at energies higher than $ \Lambda_\TC $. In the MFPC-EFT such contributions show up as NLO operators which have been described in Ref.~\cite{Cacciapaglia:2017cdi}:
	\begin{align}
	S_\mathrm{UV} =& \dfrac{s_\theta^2 C_{WW} }{\pi }\, ,\\
	T_\mathrm{UV} =& \dfrac{ s_\theta^2 (C_{\Pi D}^1 +C_{\Pi D}^2)}{16 \pi c_w^2}
	+ \dfrac{s_\theta^2 (C_{y \Pi D}^{1} + C_{y \Pi D}^4)}{64 \pi^2 \alpha} \left( 3\Tr[X_{\ubar \ubar} - X_{\dbar \dbar}] - \Tr[X_{\ebar \ebar}] \right)^2 \nonumber \\
	& \; -\dfrac{s_\theta^2 (C_{y \Pi D}^{3} + C_{y \Pi D}^6)}{64 \pi^2 \alpha} \Tr\left[3 (X_{\ubar \ubar} X_{\ubar \ubar} - 2 X_{\ubar \dbar} X_{\dbar \ubar} + X_{\dbar \dbar} X_{\dbar \dbar}) + X_{\ebar \ebar} X_{\ebar \ebar} \right] \, .
	\end{align}
The strong coefficients appearing in these contributions are the coefficients of the relevant NLO corrections to the kinetic terms (the terms have been included in appendix \ref{app:NLOoperators} for completeness).
Combining the contributions from the changed Higgs sector and those coming from UV physics through new effective operators, the total deviation from the SM prediction of the oblique parameters are
	\begin{equation}
	S = S_\mathrm{UV} + S_\mathrm{IR} \andeq T = T_\mathrm{UV} + T_\mathrm{IR}. \end{equation}
The uncertainty in the strong coefficients will make it difficult to make a true prediction as to the $ S $ and $ T $ parameters. Since in addition these coefficients are independent of the ones appearing in the
flavor observables that are in the focus of the present study, we will not consider them in our numerical analysis.

\subsection{Low-energy probes of flavor and CP violation}\label{sec:observables}

Precision measurements of flavor-changing neutral current (FCNC) processes like
meson-antimeson mixing and rare decays of $K$ and $B$ mesons are well known
to be important constraints on models with new strong dynamics.
But also flavor-changing \textit{charged} currents, mediated by the $W$
boson at tree level in the SM, are relevant since models with partial compositeness
can violate lepton flavor universality or the unitarity of the CKM matrix.
We use the open source package \texttt{flavio} \cite{flavio} for our numerics.

\subsubsection{Meson-antimeson mixing}\label{sec:df2}

The part of the weak effective Hamiltonian responsible for meson-antimeson
mixing in the $K^0$, $B^0$, and $B_s$ systems reads
\begin{equation}
\mathcal{H}_\mathrm{weak}^{\Delta F=2} = - \sum_i C_{i} O_{i} \,,
\end{equation}
where the sum runs over the following operators,
\begin{align}
O_{VLL}^{ij} &= (\bar d^j_L \gamma^\mu d^i_L)(\bar d^j_L \gamma_\mu d^i_L)\,,
&
O_{VRR}^{ij} &= (\bar d^j_R \gamma^\mu d^i_R)(\bar d^j_R \gamma_\mu d^i_R)\,,
&
O_{VLR}^{ij} &= (\bar d^j_L \gamma^\mu d^i_L)(\bar d^j_R \gamma_\mu d^i_R) \,,
\nonumber \\
O_{SLL}^{ij} &= (\bar d^j_R  d^i_L)(\bar d^j_R  d^i_L)\,,
&
O_{SRR}^{ij} &= (\bar d^j_L d^i_R)(\bar d^j_L  d^i_R)\,,
&
O_{SLR}^{ij} &= (\bar d^j_R  d^i_L)(\bar d^j_L  d^i_R)\,,
\\
O_{TLL}^{ij} &= (\bar d^j_R \sigma^{\mu\nu} d^i_L)(\bar d^j_R \sigma_{\mu\nu} d^i_L)\,,
&
O_{TRR}^{ij} &= (\bar d^j_L \sigma^{\mu\nu} d^i_R)(\bar d^j_L \sigma_{\mu\nu} d^i_R)\,, \nonumber
\end{align}
where $ij=21,31,32$ for $K^0$, $B^0$, and $B_s$, respectively. In the MFPC model,
new physics contributions to all eight operators are generated from the operators
in section~\ref{sec:effective_model}. There are two contributing mechanisms:
direct contributions from the four-fermion operators $\mathcal O^i_{4f}$ that
contain the operators in $\mathcal{H}_\mathrm{weak}^{\Delta F=2}$,
and $Z$-mediated contributions from flavor-changing $Z$ couplings induced by
the operator $\mathcal O_{\Pi f}$. In the limit of small $s_\theta$, the
latter are however subleading. To leading order\footnote{%
In our numerical analysis, we will keep also subleading terms.}
in $s_\theta$, only four
operators are generated,
\begin{align}
C_{VLL}^{ij} &=
 \big(X_{dd}^*\big)_{ij}\,
 \big(X_{dd}^*\big)_{ij}\,
 \frac{ C^4_{4f}+C^5_{4f} }{\Lambda_\text{TC}^2},
\\
C_{VRR}^{ij} &=
 \big(X_{\bar{d}\bar{d}}\big)_{ij}\,
 \big(X_{\bar{d}\bar{d}}\big)_{ij}\,
 \frac{ C^4_{4f}+C^5_{4f} }{\Lambda_\text{TC}^2},
\\
C_{VLR}^{ij} &=
 \big(X_{dd}^*\big)_{ij}\,
 \big(X_{\bar{d}\bar{d}}\big)_{ij}\,
 \frac{ C^4_{4f} }{\Lambda_\text{TC}^2},
\\
C_{SLR}^{ij} &=
 \big(Y_{d\bar{d}}\big)_{ij}\,
 \big(Y_{\bar{d}d}^*\big)_{ij}\,
 \frac{ C^2_{4f} }{\Lambda_\text{TC}^2}.
\end{align}
The combination of fundamental Yukawa couplings in $C_{SLR}^{ij}$ turns out
to be proportional to the square of the down quark mass matrix, which is
diagonal in the mass basis by definition. Thus, the operator $O_{SLR}$ is
flavor-diagonal and does \textit{not} contribute to meson-antimeson mixing.
The vanishing of this Wilson coefficient at leading order in $s_\theta$
is in contrast to effective models of partial compositeness or
extra-dimensional models based on flavor anarchy and is a consequence of our
assumption of a flavor-trivial mass matrix for the elementary scalars.
However, even for a vanishing $C_{SLR}$ at the \textit{electroweak} scale
-- which is where we match the MFPC-EFT onto the WEH --
the QCD renormalization group (RG) running down to the hadronic scale of the order
of a few GeV induces a sizable contribution to $C_{SLR}$ proportional to $C_{VLR}$.

The two left-right operators are well-known to be most problematic in models based
on partial compositeness, in particular in the kaon sector where their QCD matrix
elements are strongly chirally enhanced in addition to the RG enhancement of the
Wilson coefficients. We thus expect the strongest bound from meson-antimeson
mixing observables to come from $\epsilon_K$, measuring indirect CP violation in
$K^0$-$\bar K^0$ mixing. Although the Wilson coefficients $C^4_{4f}$ and
$C^5_{4f}$ are real, a sizable CP-violating phase in the mixing amplitude
can be induced by the fundamental Yukawa couplings.

\subsubsection{Rare semi-leptonic \texorpdfstring{$B$}{B} decays}\label{sec:fcnc}

Decays based on the $b\to s\ell\ell$ transition, such as $B\to K^*\ell\ell$
or $B\to K\ell\ell$ with $\ell=e$ or $\mu$, are probes of flavor violation
that are complementary to meson-antimeson mixing. On the one hand,
since they only involve
one flavor change, they are much more sensitive to contributions mediated
by flavor-changing $Z$ couplings induced by $\mathcal O_{\Pi f}$.
On the other, recent hints for violation of lepton flavor universality (LFU)
between the electronic and muonic $B\to K^*\ell\ell$ and $B\to K\ell\ell$ rates
raise the question whether -- and to what level -- LFU can be violated in MFPC.
To leading order in $s_\theta$, the $Z$-mediated contributions are lepton
flavor universal, but direct contributions from the four-fermion
operators $\mathcal O^i_{4f}$ containing two quarks and two leptons are in fact
\textit{expected} to violate LFU and enter at the same order in $s_\theta$
as the $Z$-mediated effects.

The effective Hamiltonian for $b\to s\ell\ell$ transitions can be written as
\begin{equation}
\mathcal{H}_\mathrm{weak}^{b\to s\ell\ell}= -
\sum_{i,\ell}
(C_i^\ell O_i^\ell + C_i^{\prime\ell} O_i^{\prime\ell}) + \text{h.c.}
\end{equation}
The most important operators for our discussion\footnote{%
In particular, we neglect dipole operators \cite{Konig:2014iqa}, which always conserve LFU.
Scalar operators are
flavor-diagonal in the mass basis and thus do not contribute.}
read
\begin{align}
O_9^\ell &=
(\bar{s}_L \gamma_{\mu} b_L)(\bar{\ell} \gamma^\mu \ell)\,,
&
O_9^{\prime\ell} &=
(\bar{s}_R \gamma_{\mu} b_R)(\bar{\ell} \gamma^\mu \ell)\,,
\\
O_{10}^\ell &=
(\bar{s}_L \gamma_{\mu} b_L)( \bar{\ell} \gamma^\mu \gamma_5 \ell)\,,
&
O_{10}^{\prime\ell} &=
(\bar{s}_R \gamma_{\mu}  b_R)( \bar{\ell} \gamma^\mu \gamma_5 \ell)\,.
\end{align}
The direct four-fermion contributions to their Wilson coefficients,
to leading order\footnote{%
In our numerical analysis, we will keep also subleading terms.}
in $s_\theta$, reads
\begin{align}
C_{9}^\ell &\supset
-
\frac{1}{4}\,
\big(X_{dd}^*\big)_{bs}\,
\big(X_{\bar{e}\bar{e}}\big)_{\ell\ell}\,
\frac{ C^4_{4f}}{\Lambda_\text{TC}^2}
+
\frac{1}{4}\,
\big(X_{dd}^*\big)_{bs}\,
\big(X_{ee}\big)_{\ell\ell}\,
\frac{ C^4_{4f}+C^5_{4f} }{\Lambda_\text{TC}^2}
\,,\\
C_{9}^{\prime\ell} &\supset
-
\frac{1}{4}\,
\big(X_{\dbar \dbar}\big)_{bs}\,
\big(X_{ee}\big)_{\ell\ell}\,
\frac{ C^4_{4f}}{\Lambda_\text{TC}^2}
+
\frac{1}{4}\,
\big(X_{\dbar \dbar}\big)_{bs}\,
\big(X_{\ebar \ebar}\big)_{\ell\ell}\,
\frac{ C^4_{4f}+C^5_{4f} }{\Lambda_\text{TC}^2}
\,,\\
C_{10}^\ell &\supset
-
\frac{1}{4}\,
\big(X_{dd}^*\big)_{bs}\,
\big(X_{\bar{e}\bar{e}}\big)_{\ell\ell}\,
\frac{ C^4_{4f}}{\Lambda_\text{TC}^2}
-
\frac{1}{4}\,
\big(X_{dd}^*\big)_{bs}\,
\big(X_{ee}\big)_{\ell\ell}\,
\frac{ C^4_{4f}+C^5_{4f} }{\Lambda_\text{TC}^2}
\,,\\
C_{10}^{\prime\ell} &\supset
+
\frac{1}{4}\,
\big(X_{\dbar \dbar}\big)_{bs}\,
\big(X_{ee}\big)_{\ell\ell}\,
\frac{ C^4_{4f}}{\Lambda_\text{TC}^2}
+
\frac{1}{4}\,
\big(X_{\dbar \dbar}\big)_{bs}\,
\big(X_{\ebar \ebar}\big)_{\ell\ell}\,
\frac{ C^4_{4f}+C^5_{4f} }{\Lambda_\text{TC}^2}
\,,
\end{align}
while the $Z$-mediated contributions can be written as
\begin{align}
C_{9}^\ell &\supset
2\pi\,
\big(X_{dd}^*\big)_{bs}\,
(4\,s_w^2-1)\,
\frac{C_{\Pi f}}{\Lambda_\text{TC}^2}
\,,\\
C_{9}^{\prime\ell} &\supset
-2\pi\,
\big(X_{\dbar \dbar}\big)_{bs}\,
(4\,s_w^2-1)\,
\frac{C_{\Pi f}}{\Lambda_\text{TC}^2}
\,,\\
C_{10}^\ell &\supset
2\pi\,
\big(X_{dd}^*\big)_{bs}\,
\frac{C_{\Pi f}}{\Lambda_\text{TC}^2}
\,,\\
C_{10}^{\prime\ell} &\supset
-2\pi\,
\big(X_{\dbar \dbar}\big)_{bs}\,
\frac{C_{\Pi f}}{\Lambda_\text{TC}^2}
\,.
\end{align}

\subsubsection{Tree-level semi-leptonic decays}\label{sec:cc}

Charged-current semi-leptonic decays based on the $q\to q' \ell\nu$
transition are mediated at tree level by the $W$ boson in the SM and are
used to measure the elements of the CKM matrix without pollution by loop-induced
new physics effects. In MFPC however, these processes can receive new physics
contributions from the operators in the MFPC-EFT. Similarly to the
semi-leptonic FCNC decays, there are contributions from modified
$W$ couplings to quarks induced by $\mathcal O_{\Pi f}$ that are lepton flavor universal
to leading order in $s_\theta$, as well as direct four-fermion contributions from
$O^i_{4f}$ that are \textit{expected} to violate LFU.
In addition, in charged-current decays, $\mathcal O_{\Pi f}$ induces contributions from modified
$W$ couplings to \textit{leptons} that are also expected to violate LFU.

Decays where $\ell$ is a light lepton, i.e.\ an electron or muon,
must be taken into account as \textit{constraints} in our analysis.
They are important for two reasons: they constrain
the amount of LFU that can potentially be observed in FCNC decays with light leptons, and they
are necessary to consistently compare the CKM matrix obtained from diagonalizing
the quark mass matrices with the CKM \textit{measurements}.

In addition, we consider the \textit{semi-tauonic} decays based on the $b\to c\tau\nu$
transition.
The world averages for the ratios $R_{D^{(*)}}$ of the $B\to D^{(*)}\tau\nu$
over the $B\to D^{(*)}\ell\nu$ ($\ell=e,\mu$) branching ratios currently deviate
from the SM prediction at a combined level of $4\sigma$ \cite{Amhis:2016xyh}. Assessing whether
the MFPC model can account for these deviations is an important goal of our study.

The effective Hamiltonian for $d_i\to u_j\ell\nu$ transitions can be written as
\begin{equation}
\mathcal{H}_\mathrm{weak}^{d_i\to u_j\ell\nu} = \sum_i C_i^{\ell(\prime)} O_i^{\ell(\prime)} + \text{h.c.},
\end{equation}
where the sum runs over the following operators,
\begin{align}
O_V^{d^iu^j\ell} &= (\bar u^j_L \gamma^\mu d^i_L)(\bar \ell_L \gamma_\mu \nu_{\ell L})
\,, &
O_V^{d^iu^j\ell\prime} &= (\bar u^j_R \gamma^\mu d^i_R)(\bar \ell_L \gamma_\mu \nu_{\ell L})
\,, \\
O_S^{d^iu^j\ell} &= m_b(\bar u^j_L d^i_R)(\bar \ell_R \nu_{\ell L})
\,, &
O_S^{d^iu^j\ell\prime} &= m_b(\bar u^j_R d^i_L)(\bar \ell_R \nu_{\ell L})
\,, \\
O_T^{d^iu^j\ell} &= (\bar u^j_R \sigma^{\mu\nu} d^i_L)(\bar \ell_R \sigma_{\mu\nu}\nu_{\ell L}) \,.
\end{align}
In the SM, $C_V^{u^id^j\ell} = 4G_FV_{ij}/\sqrt{2}$ and all other coefficients
vanish. In MFPC, all of them are generated.
The direct four-fermion contributions to their Wilson coefficients,
to leading order\footnote{%
In our numerical analysis, we will keep also subleading terms.}
in $s_\theta$, reads
\begin{align}
C_{V}^{d^iu^j\ell} &\supset
\frac{1}{2}\,
\big(X_{du}^*\big)_{ij}\,
\big(X_{e\nu}\big)_{\ell\ell}\,
\frac{ C^5_{4f} - C^3_{4f} }{\Lambda_\text{TC}^2}
\,,\\
C_{V}^{d^iu^j\ell\prime} &\supset 0
\,,\\
C_{S}^{d^iu^j\ell} &\supset
\big(Y_{\dbar u}^*\big)_{ij}\,
\big(Y_{\ebar\nu}\big)_{\ell\ell}\,
\frac{ C^2_{4f} }{\Lambda_\text{TC}^2}
\,,\\
C_{S}^{d^iu^j\ell\prime} &\supset
\frac{1}{2}
\big(Y_{d \ubar}\big)_{ij}\,
\big(Y_{\ebar\nu}\big)_{\ell\ell}\,
\frac{ C^{8*}_{4f}-2\,C^{7*}_{4f} }{\Lambda_\text{TC}^2}
\,,\\
C_{T}^{d^iu^j\ell} &\supset
\frac{1}{8}
\big(Y_{d \ubar}\big)_{ij}\,
\big(Y_{\ebar\nu}\big)_{\ell\ell}\,
\frac{ C^{8*}_{4f} }{\Lambda_\text{TC}^2}
\,,
\end{align}
while the $W$-mediated contributions read
\begin{align}
C_{V}^{d^iu^j\ell} &\supset
-8\,\pi\,
\left(
\big(X_{du}^*\big)_{ij}\,
+
V_{ji}\,
\big(X_{e\nu}\big)_{\ell\ell}
\right)
\frac{ C_{\Pi f} }{\Lambda_\text{TC}^2}
\,,\\
C_{V}^{d^iu^j\ell\prime} &\supset
8\,\pi\,
\big(X_{\dbar \ubar}\big)_{ij}\,
\frac{ C_{\Pi f} }{\Lambda_\text{TC}^2}
\,.
\end{align}
As \textit{constraints}, we consider the following processes sensitive to these Wilson coefficients:
\begin{itemize}
\item For $d\to u\ell\nu$, the branching ratio of $\pi^+\to e\nu$ (which is sensitive
to $e$-$\mu$ LFU violation since the branching ratio of the muonic mode is almost 100\%),
\item For $s\to u\ell\nu$, the branching ratio of $K^+\to \mu\nu$ and the
 ratio of $K^+\to \ell\nu$ branching ratios with $\ell=e$ and $\mu$,
\item For $b\to c\ell\nu$, the branching ratios of $B\to D\ell\nu$ and $B\to D^*\ell\nu$ with $\ell=e$ and $\mu$.
\end{itemize}
As \textit{predictions}, we further consider:
\begin{itemize}
\item For $b\to c\tau\nu$, the ratios $R_D$ and $R_{D^*}$.
\end{itemize}
Table~\ref{tab:exp} lists all the experimental values and SM predictions according to \texttt{flavio} v0.23 used in our analysis. Note that the uncertainties on the SM prediction shown in this table include (and in
many cases are dominated by) the parametric uncertainties due to the limited knowledge
of CKM elements. In our numerical scan, as detailed in the following section,
CKM parameters are predicted as functions of the model parameters, such that only
the non-CKM uncertainties are relevant for the $\chi^2$ in any given parameter point.

\begin{table}
\renewcommand{\arraystretch}{1.3}
\begin{tabular}{llll}
\hline
Observable & measurement & & SM prediction \\
\hline
$\Delta M_s$ & $(17.76 \pm 0.02) $ ps & \cite{Amhis:2014hma} & $(19.9\pm1.7)$ ps \\
$\Delta M_d$ & $(0.505 \pm 0.002)$ ps & \cite{Amhis:2014hma} & $(0.64\pm0.09)$ ps \\
$S_{\psi\phi}$ & $(3.3 \pm 3.3) \times 10^{-2}$ & \cite{Amhis:2014hma} & $(3.75\pm0.22) \times 10^{-2}$ \\
$S_{\psi K_S}$ & $0.679 \pm 0.020$ & \cite{Amhis:2014hma} & $0.690\pm0.025$ \\
$\vert\epsilon_K\vert$ & $(2.228 \pm 0.011) \times 10^{-3}$ & \cite{Agashe:2014kda} & $(1.76\pm0.22) \times 10^{-3}$ \\
\hline
$\text{BR}(B^+\to D^0\ell^+\nu_\ell)$ & $(2.330 \pm 0.098) \times 10^{-2}$ & \cite{Amhis:2014hma} & $(2.92\pm0.21) \times 10^{-2}$ \\
$\text{BR}(B^0\to D^{\ast -}\ell^+\nu_\ell)$ & $(4.88 \pm 0.10) \times 10^{-2}$ & \cite{Amhis:2014hma} & $(5.72\pm0.27) \times 10^{-2}$ \\
$\text{BR}(\pi^+\to e^+\nu)$ & $(1.234\pm0.002) \times 10^{-4}$ & \cite{Aguilar-Arevalo:2015cdf} & $(1.2341\pm0.0002) \times 10^{-4}$ \\
$\text{BR}(K^+\to \mu^+\nu)$ & $0.6356 \pm 0.0011$ & \cite{Agashe:2014kda} & $0.6296\pm0.0066$ \\
$R_{e\mu}(K^+\to \ell^+\nu)$ & $(2.488 \pm 0.009) \times 10^{-5}$ & \cite{Agashe:2014kda} & $(2.475\pm0.001) \times 10^{-5}$ \\
\hline
$R_{D}$ & $0.397\pm0.049$ & \cite{Amhis:2016xyh} & $0.277\pm0.012$ \\
$R_{D^*}$ & $0.316\pm0.019$ & \cite{Amhis:2016xyh} & $0.2512\pm0.0043$ \\
$R_K^{[1, 6]}$ & $0.75 ^{+ 0.08}_{ - 0.10}$ & \cite{Aaij:2014ora} & $1.000\pm0.001$ \\
$R_{K^*}^{[0.045, 1.1]}$ & $0.65 ^{+ 0.07}_{ - 0.12}$ & \cite{Aaij:2017vbb} & $0.926 \pm 0.004$ \\
$R_{K^*}^{[1.1, 6.0]}$ & $0.68 ^{+ 0.08}_{ - 0.12}$ & \cite{Aaij:2017vbb} & $0.9965 \pm 0.0005$  \\
\hline
\end{tabular}
\caption{Measurements and SM predictions (computed with \texttt{flavio} v0.23)
of flavor observables used in our analysis. The first two blocks are the
meson-antimeson mixing and charged current observables used as \textit{constraints},
while the observables in the last block are considered as \textit{predictions}.}
\label{tab:exp}
\end{table}

\section{Numerical analysis}\label{sec:numerics}

To investigate possible NP effects of the MFPC model on the low-energy observables discussed above, we calculate predictions for these observables, depending on the position in the parameter space of the MFPC-EFT.
To avoid strong constraints from charged lepton flavor violation (see e.g.\ \cite{KerenZur:2012fr}),
we assume that the fundamental Yukawa coupling matrices $y_L$ and $y_{\bar e}$ can be diagonalized
in the same basis at the matching scale\footnote{Note that this assumption is not
renormalization group invariant in the presence of lepton flavor universality
violation \cite{Feruglio:2017rjo}.}.
\subsection{Parameters}\label{sec:numerics_parameters}
The observables in our analysis depend on the following MFPC-EFT parameters:
\begin{itemize}
 \item The new strong coupling scale $\Lambda_\TC=4\pi f_\TC$. We vary $f_\TC$ between 1~TeV and 3~TeV.
 \item The six real Wilson coefficients $C^{1}_{4f}$, $C^{2}_{4f}$, $C^{3}_{4f}$, $C^{4}_{4f}$, $C^{5}_{4f}$ and $C_{\Pi f}$. We vary their absolute values logarithmically between $0.1$ and $10$ and allow them to have either sign.
 \item The four complex Wilson coefficients $C^{6}_{4f}$, $C^{7}_{4f}$, $C^{8}_{4f}$ and $C_{\rm Yuk}$. We vary their absolute values logarithmically between $0.1$ and $10$ as well as their complex phases linearly between $0$ and $2\pi$.
 \item The four\footnote{Since we assume $y_\ebar$ to be diagonal in the same basis as $y_L$, its entries are fixed by requiring that the product of $y_L$ and $y_\ebar$ yields the correct masses for the charged leptons.} fundamental Yukawa coupling matrices $y_Q$, $y_L$, $y_\ubar$, $y_\dbar$.
 For parameterizing them, we first introduce the effective Yukawa matrices
 \begin{equation}
  \tilde{y}_f = \sqrt{C_{\rm Yuk}}\,y_f,
 \end{equation}
 which allow for expressing the SM fermion mass matrices independently of $C_{\rm Yuk}$.
 Each complex matrix $\tilde{y}_f$ can in general be written in terms of one positive real diagonal and two unitary matrices.
 One of those two unitary matrices can always be absorbed in a redefinition of the SM fields.
 For two of the matrices $\tilde{y}_f$, the second unitary matrix can be absorbed into the techniscalar fields, and thus two effective Yukawa matrices can be chosen to be positive real diagonal. We choose
 \begin{equation}
  \tilde{y}_Q = {\rm diag}(y_{Q1},y_{Q2},y_{Q3}),
  \quad\quad
  \tilde{y}_L = {\rm diag}(y_{L1},y_{L2},y_{L3}).
 \end{equation}
 Parameterizing the two remaining unitary matrices that enter $\tilde{y}_\ubar$, $\tilde{y}_\dbar$ by in total six angles $t_{u}^{12}$ ,$t_{u}^{13}$, $t_{u}^{23}$, $t_{d}^{12}$ ,$t_{d}^{13}$, $t_{d}^{23}$ and four phases\footnote{A general $3\times3$ unitary matrix has five independent phases. However, six of the phases of $\tilde{y}_\ubar$ and $\tilde{y}_\dbar$ can be absorbed by field redefinitions, leaving four phases in total.} $\delta_d$, $\delta_u$, $a_d$, $b_d$, we get
 \begin{equation}
 \begin{aligned}
  \tilde{y}_\ubar &= {\rm unitary}(t_{u}^{12},t_{u}^{13},t_{u}^{23},\delta_u)\cdot{\rm diag}(y_{u1},y_{u2},y_{u3}),
  \\
  \tilde{y}_\dbar &= {\rm unitary}(t_{d}^{12},t_{d}^{13},t_{d}^{23},\delta_d,a_d,b_d)\cdot{\rm diag}(y_{d1},y_{d2},y_{d3}).
 \end{aligned}
 \end{equation}
 We vary the diagonal entries logarithmically between\footnote{The lower boundary for the diagonal entries of $\tilde{y}_L$ is adjusted such that the diagonal entries of $y_\ebar$ stay below $4\pi$ when requiring the correct charged lepton masses.} $0.002$ and $4\pi$ and the angles and phases linearly between $0$ and $2\pi$.
\end{itemize}
We thus have in total 14 real parameters for the Wilson coefficients, 22 real parameters for the fundamental Yukawa matrices and one real parameter for the new strong scale.
The Wilson coefficients as well as the fundamental Yukawa matrices are defined at the matching scale, i.e at 160~GeV.

\subsection{Strategy}

Given the high dimensionality of the parameter space, a naive brute-force scan by randomly choosing each of the parameters is not applicable.
We observe, however, that the quark masses and CKM elements only depend on the effective Yukawa matrices $\tilde{y}_Q$, $\tilde{y}_\ubar$ and $\tilde{y}_\dbar$ (see section \ref{sec:sm_yukawas}).
This can be used in a first step to find a region in parameter space where the predictions for the quark masses and CKM elements are close to experimental observations.
In this step, only the effective quark Yukawa matrices have to be varied.
The lepton Yukawa matrix $\tilde{y}_L$, all MFPC-EFT Wilson coefficients and the new strong scale do not enter.
In a second step, one can then randomly choose the remaining parameters while preserving the predictions of SM fermion masses\footnote{As described above, by adjusting $\tilde{y}_\ebar$, the charged lepton masses are always fixed to their experimental value and are thus unaffected by varying $\tilde{y}_L$.} and CKM elements.

For predicting the quark masses, we construct the mass matrix in eq.~(\ref{eq:SM_masses}) from the effective Yukawa matrices and numerically diagonalize it via eq. (\ref{eq:mass_basis}).
We interpret each quark mass as $\overline{\rm MS}$ running mass at 160 GeV and run it to the scale where it can be compared to its PDG average.
The numerical diagonalization also yields the rotation matrices from which we calculate the CKM elements via eq. (\ref{eq:CKM}).
However, the CKM elements cannot be directly compared to the experimental values, as the observables are affected by dimension-six operators too. 
Contrary to the quark masses, we consequently cannot impose the constraints on the CKM elements already in the first step of the scan.
The CKM elements are therefore constrained in the second step by the charged-current semi-leptonic decays discussed in section~\ref{sec:cc}.
This is done after taking the contributions from dimension-six operators into account.
In the first step, however, we require the CKM elements to be close to certain input values that we have found to yield many points that pass the constraints imposed in the second step.
To compare the predictions for the masses to their PDG averages and the predictions for the CKM elements to our input values, we construct a $\chi^2$-function $\chi^2_\text{mass, CKM}$.
This function only depends on the 19 parameters of $\tilde{y}_Q$, $\tilde{y}_\ubar$ and $\tilde{y}_\dbar$.
We then proceed in the following way:
\begin{itemize}
 \item Starting from a randomly chosen point in the 19-dimensional parameter-subspace where $\chi^2_\text{mass, CKM}$ lives, we numerically minimize $\chi^2_\text{mass, CKM}$ to find a viable point that predicts correct quark masses and CKM elements close to our input values.
 \item Starting from this viable point, we use a Markov-Chain for an efficient sampling of the parameter space, as first proposed in \cite{Straub:2013zca} and also applied in \cite{Niehoff:2015iaa,Niehoff:2016zso}.
 This is done by employing the Markov-Chain-Monte-Carlo implementation from the \texttt{pypmc} package \cite{pypmc}.
 The chain samples the region around the previously found minimum and generates 10\,k viable points with a low value of $\chi^2_\text{mass, CKM}$.
 \item We reduce the auto-correlation of the 10\,k viable points generated by the Markov-Chain by selecting only 1000 points.
\end{itemize}
The above steps are repeated 100\,k times to yield 100\,M points from 100\,k local minima of $\chi^2_\text{mass, CKM}$ that all predict CKM elements close to our input values and correct quark masses.

For these points we then randomly choose the remaining 18 parameters and calculate all the observables discussed in sections~\ref{sec:observables} and \ref{sec:constraints_Z} using the open source package \texttt{flavio} \cite{flavio}.
We subsequently construct $\chi^2$-functions for three classes of constraints:
\begin{itemize}
 \item $\chi^2_\text{$Z$}$ compares the experimental values shown in table~\ref{tab:experimental_values} to our predictions of $Z$-decay observables discussed in section~\ref{sec:constraints_Z}.
 \item $\chi^2_{\Delta F=2}$ compares the meson-antimeson mixing constraints from table~\ref{tab:exp} to the predictions of the observables discussed in section~\ref{sec:df2}.
 \item $\chi^2_\text{CC}$ compares the constraints from semi-leptonic charged-current decays from table~\ref{tab:exp} to the predictions of the observables discussed in section~\ref{sec:cc}.
\end{itemize}
These $\chi^2$ functions are then used to apply the various experimental constraints on the parameter points.

\subsection{Results}\label{sec:results}

\subsubsection{Meson-antimeson mixing}

\begin{figure}
\includegraphics[width=10cm]{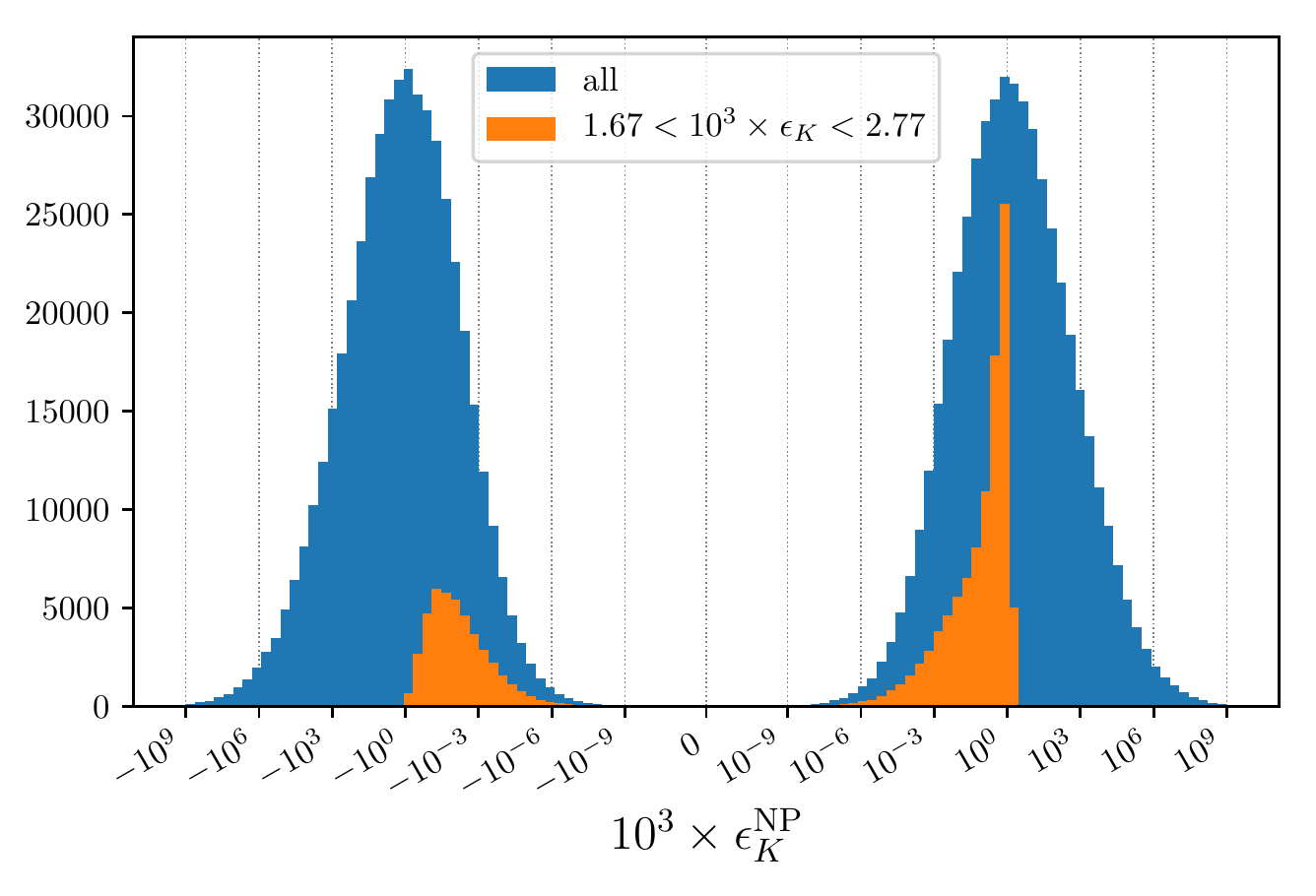}
\caption{Histogram showing the NP contribution to $\epsilon_K$ for a representative subset of all
points that feature the right masses and CKM elements,
compared to the points among those that pass the experimental constraint.
A positive NP contribution corresponds to constructive interference with the SM.}
\label{fig:epsKhist}
\end{figure}

As discussed in section~\ref{sec:df2}, the constraints from meson-antimeson
mixing, in particular the neutral kaon sector, are expected to be very important
in case of ``flavor anarchic'' fundamental Yukawa couplings. This is confirmed
by our numerical findings, where many parameter points that have the correct
quark masses and CKM mixing angles predict an order-of-magnitude enhancement
of $\epsilon_K$.
This ``$\epsilon_K$ problem'', that plagues all models with partial compositeness
(or its extra-dimensional dual description) without additional flavor symmetries
\cite{Csaki:2008zd,Blanke:2008zb,Bauer:2009cf},
is often phrased as requiring a scale $\Lambda_\text{TC}$ in excess of 15~TeV,
based on a naive estimate
$C_{VLR}\sim C_{SLR} \sim m_d m_s/(v^2\Lambda_\text{TC}^2)$.
However, the exact result depends strongly on the precise form of the fundamental
Yukawa couplings and can deviate from this naive estimate by orders of magnitude
in either direction.
In fact, we do find a significant number of points where $\epsilon_K$
is within the experimentally allowed range.
To get a feeling of the size of the new physics contributions to $\epsilon_K$\footnote{%
Here we are referring to the genuine dimension-6 NP contributions but remind the reader
that CKM elements are varied during our scan, so also the SM prediction itself
differs from point to point.},
we present the histogram in fig.~\ref{fig:epsKhist}.
It includes a representative subset of all the points that have the correct fermion masses and
CKM matrix, along with the points surviving the $\epsilon_K$ constraint\footnote{%
An interesting feature of the histogram is the fact that there are more allowed points
with a NP contribution to $\epsilon_K$ interfering constructively with the SM.
The reason is that, as discussed above, we used the exclusive semi-leptonic decays $B\to D^{(*)}\ell\nu$
as constraints in our scan. They currently prefer a lower value of $V_{cb}$ compared
to the inclusive semi-leptonic decay. Since the SM prediction of $\epsilon_K$
is highly sensitive to the value of $V_{cb}$, this tends to lead to a value
that is on the low side of the measurement \cite{Brod:2010mj}, favoring constructively interfering NP.}.  
This histogram shows that the new physics contribution varies over many orders
of magnitude. Our variation of the Wilson coefficients,
which enter linearly,
between $0.1$ and $10$ is only partially responsible for this variation.

\begin{figure}
\includegraphics[width=\textwidth]{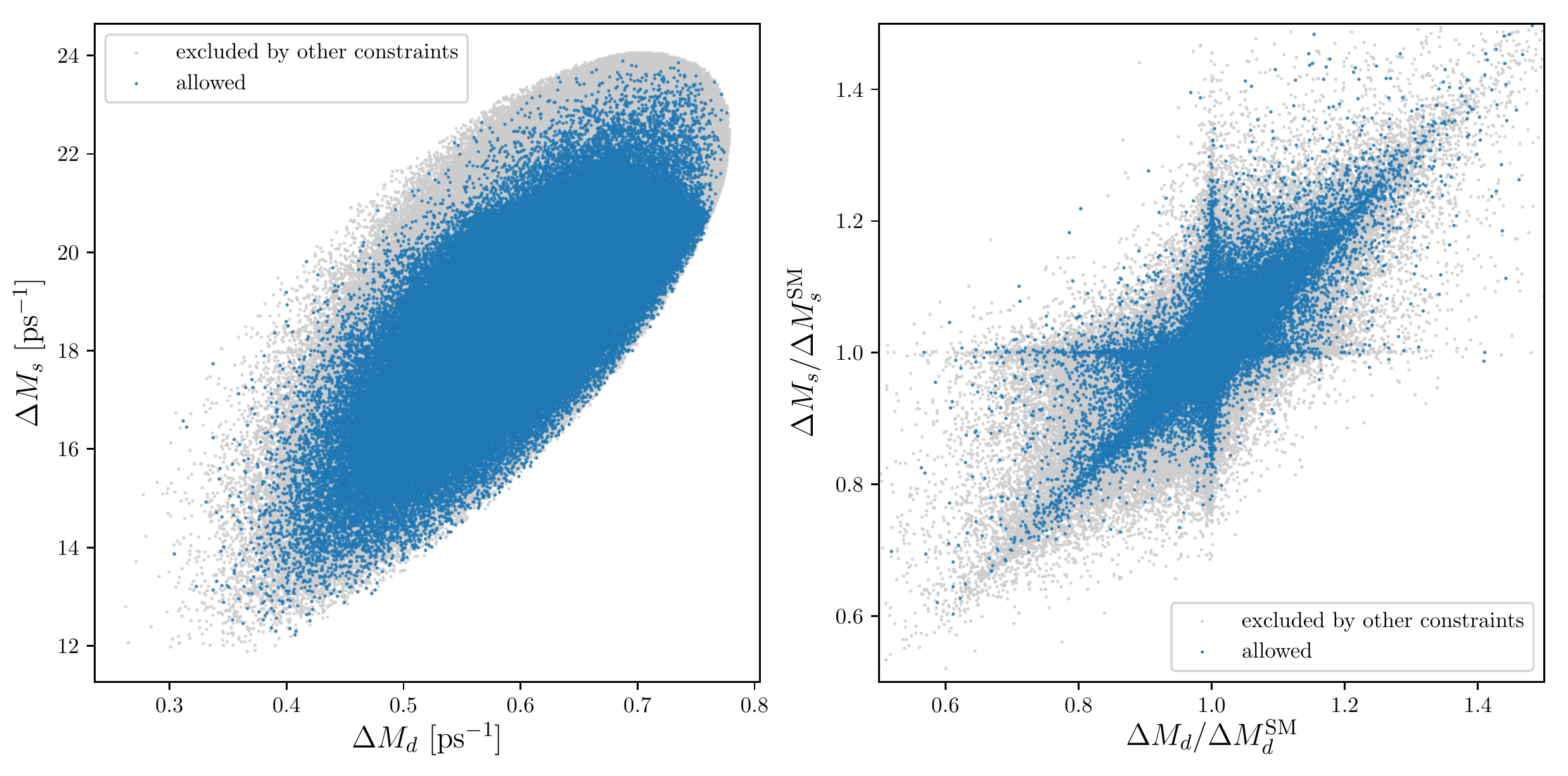}
\caption{Predictions for $\Delta M_d$ and $\Delta M_s$.
Gray points are excluded by constraints other than $\Delta F=2$.
Blue points are allowed by all constraints.}
\label{fig:DMq}
\end{figure}

New physics contributions to $B^0$ and $B_s$ meson mixing are generated as well, even though the effects are less problematic than in $K^0$ mixing since
the chiral enhancement of the $LR$ operators is absent.
In figure~\ref{fig:DMq}, we show the predictions for the mass differences
$\Delta M_d$ and $\Delta M_s$ for all our allowed points as well as for the points excluded by constraints other than meson-antimeson mixing. We emphasize again that the CKM parameters are varied during our scan. Consequently, the allowed ranges for $\Delta M_d$ and $\Delta M_s$ for a given parameter point, with fixed CKM elements, are determined by the experimental measurements smeared by the uncertainties of the matrix elements from lattice QCD \cite{Bazavov:2016nty}.
The elliptic outline visible in the left-hand panel of figure~\ref{fig:DMq}
corresponds to these allowed ranges imposed at $3\sigma$ in our scan.
The reason for the allowed (blue) points clustering in the lower part of this ellipse is that the maximal values of $\Delta M_s$ are most easily accessed
for high values of $V_{cb}$, that are however disfavored by the $B\to D\ell\nu$ branching ratio imposed in our scan. To disentangle the shifts in
$\Delta M_d$ and $\Delta M_s$ due to variation of CKM parameter vs.\ genuine dimension-6 new physics contributions, it is instructive to plot the total contribution divided by the SM contribution for the given value of the CKM parameters at each point. The result is shown in the right-hand panel of
figure~\ref{fig:DMq}. The allowed points show relative modifications of both
observables of up to 40\% with respect to the SM; this is possible since the
modifications can be partially compensated by shifts in the CKM parameters.
Both observables can be enhanced or suppressed. We further observe three clusters of points with sizable new physics effects: where mostly
$\Delta M_d$ is affected, where mostly $\Delta M_s$ is affected, and where both are affected in the same way.

\begin{figure}
\includegraphics[width=\textwidth]{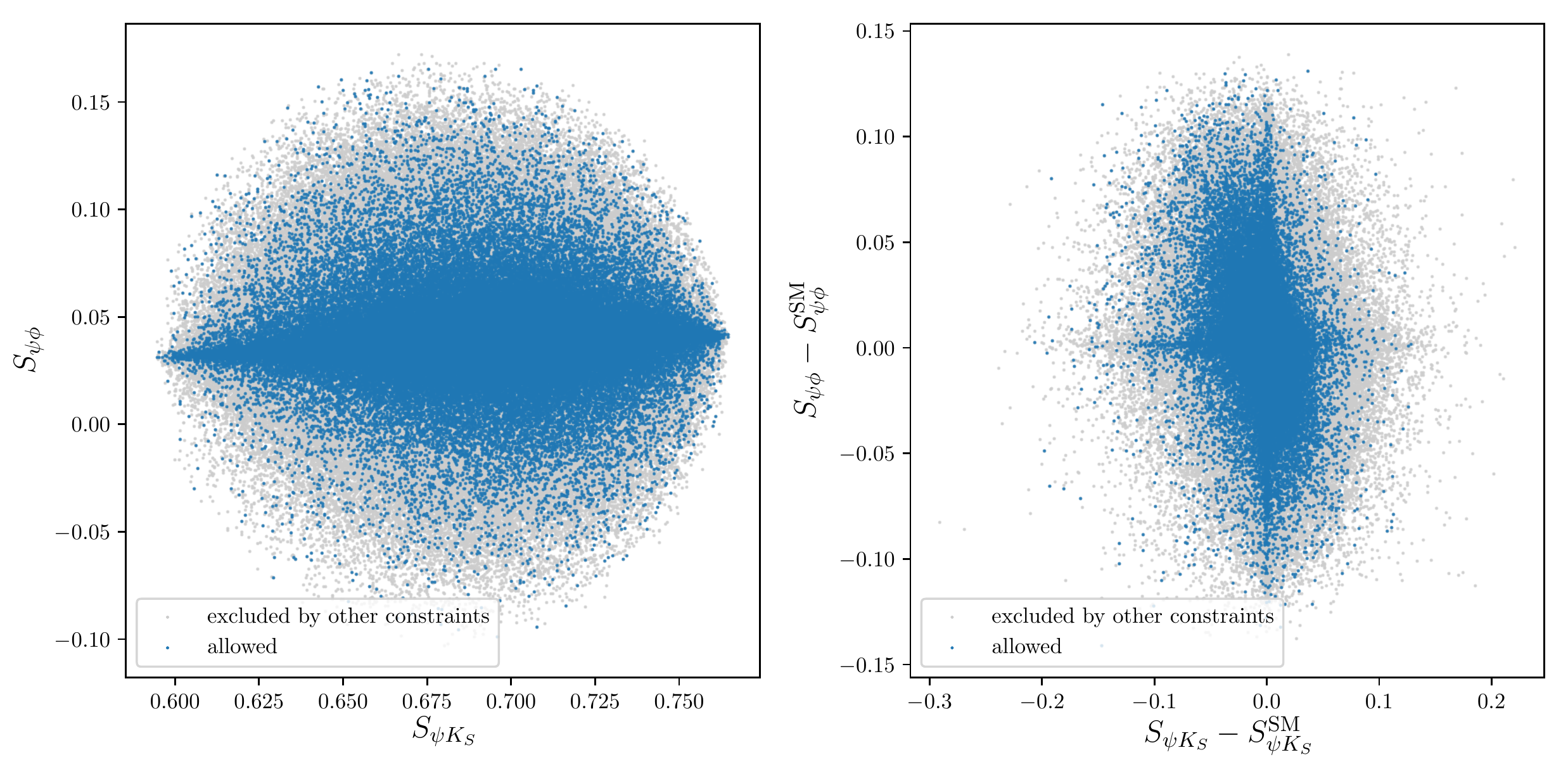}
\caption{Predictions for the mixing induced CP asymmetries
in $B^0\to J/\psi K_S$ and $B_s\to J/\psi\phi$,
sensitive to the $B^0$ and $B_s$ mixing phases.
Gray points are excluded by constraints other than $\Delta F=2$.
Blue points are allowed by all constraints.}
\label{fig:SpsiX}
\end{figure}

Apart from modifying the mass differences in the $B^0$ and $B_s$ systems,
also new CP-violating phases can be generated in the mixing amplitudes.
These can be probed in the mixing induced CP asymmetries
in $B^0\to J/\psi K_S$ and $B_s\to J/\psi\phi$.
The predictions for these observables are shown in figure~\ref{fig:SpsiX}.
The left-hand panel again shows the allowed points due to variation of CKM
elements and new physics contributions, while the right-hand
panel shows the shift in the asymmetries due to genuine dimension-6
new physics contribution by subtracting the SM contribution for the
given values of CKM elements in each point. We observe that the shift in
both asymmetries can be of order $0.1$ and we again observe clusters of points
with sizable effects where mostly one of the two observables is affected.

\subsubsection{Tree-level decays and lepton flavor universality}

\begin{figure}
\includegraphics[width=\textwidth]{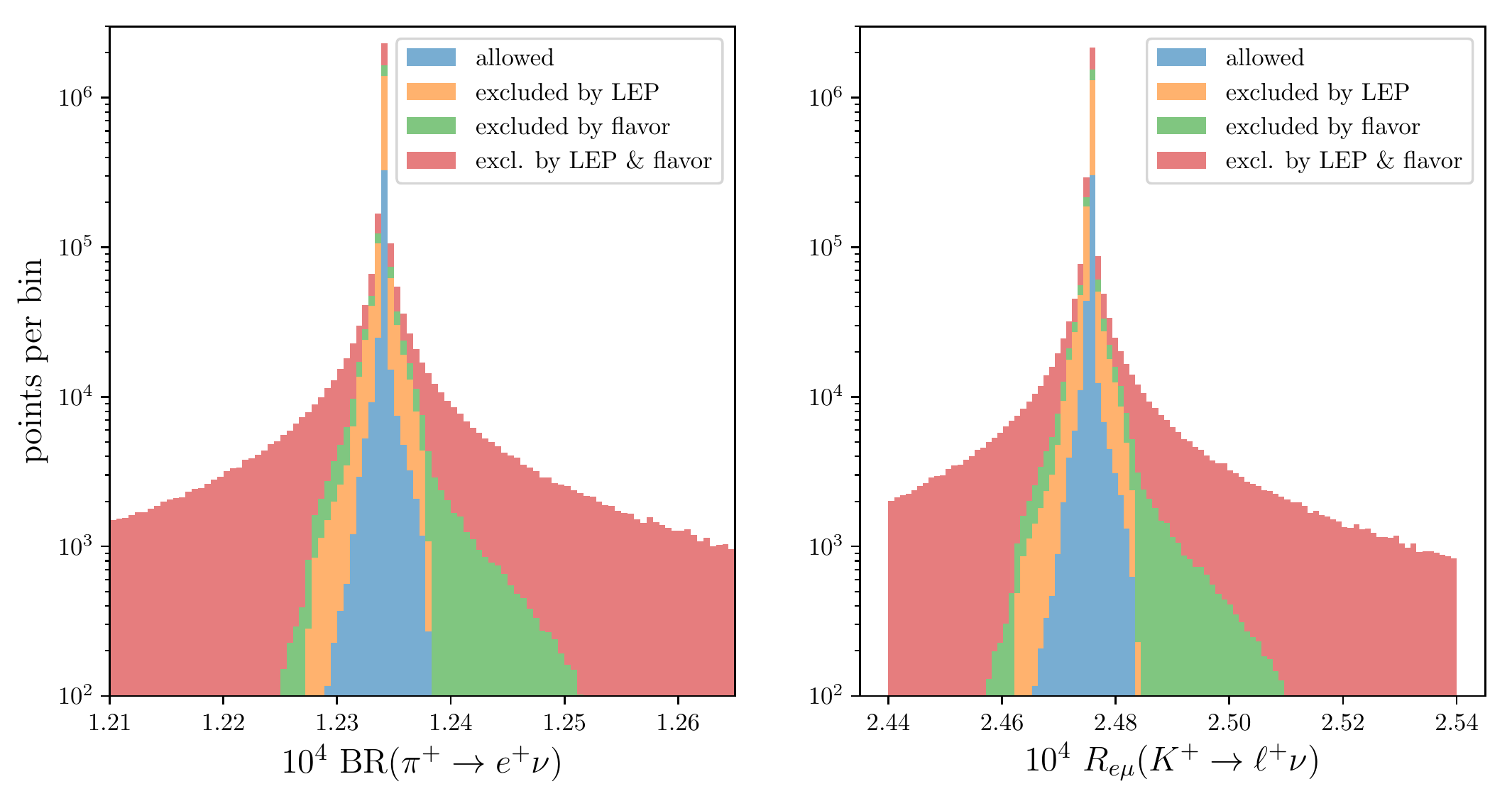}
\caption{Histogram showing the distribution of the predictions for
two observables probing $e$-$\mu$ universality violation in $Z$ couplings
for all points passing the meson-antimeson mixing constraints.
Points labeled ``excluded by LEP'' are excluded by the partial $Z$ width
measurements at LEP, while points labeled ``excluded by flavor''
are excluded by one of the charged-current decays imposed as constraints.}
\label{fig:pienu_Klnu}
\end{figure}

The precise measurements of $\text{BR}(\pi\to e\nu)$ and
$R_{e\mu}(B\to K\ell\nu)=\text{BR}(K\to e\nu)/\text{BR}(K\to \mu\nu)$,
that we impose as constraints in our analysis, lead to a strong restriction
of $e$-$\mu$ universality violation.
This is important since we are interested in the allowed size of
$e$-$\mu$ universality violation in flavor-changing \textit{neutral}
currents, as indicated by LHCb measurements.
In our scan, we find points where the deviations in these two observables
are much larger than allowed by experiments, but we find the \textit{ratio}
of the two to always be SM-like. This can be easily understood since the
dominant effects in these transitions involving light quarks, $u\to d\ell\nu$
or $s\to u\ell\nu$, is through a modified $W$ coupling to leptons induced
by the operator $\mathcal O_{\Pi f}$, while the direct four-fermion contributions induced by the operators $\mathcal O_{4f}^i$ are suppressed
by the small fundamental Yukawa couplings of the light quark generations.
By $\SU(2)_\LL$ symmetry,
this lepton flavor non-universal modification of $W$ couplings implies
a corresponding modification of $Z$ couplings that is constrained by $Z$ pole measurements at LEP. In figure~\ref{fig:pienu_Klnu},
we show a histogram of the values for the two observables of interest
for all the points passing the meson-antimeson mixing constraints.
We distinguish points excluded by LEP, excluded by flavor (i.e.\ one of the
\textit{charged-current} decays imposed as constraints in the analysis),
excluded by both, and allowed by all constraints. These plots demonstrate
that LEP and flavor constraints are both relevant to constrain
$e$-$\mu$ universality violation in $Z$ couplings and that the resulting
constraint is at the per cent level.

\begin{figure}
\includegraphics[width=0.55\textwidth]{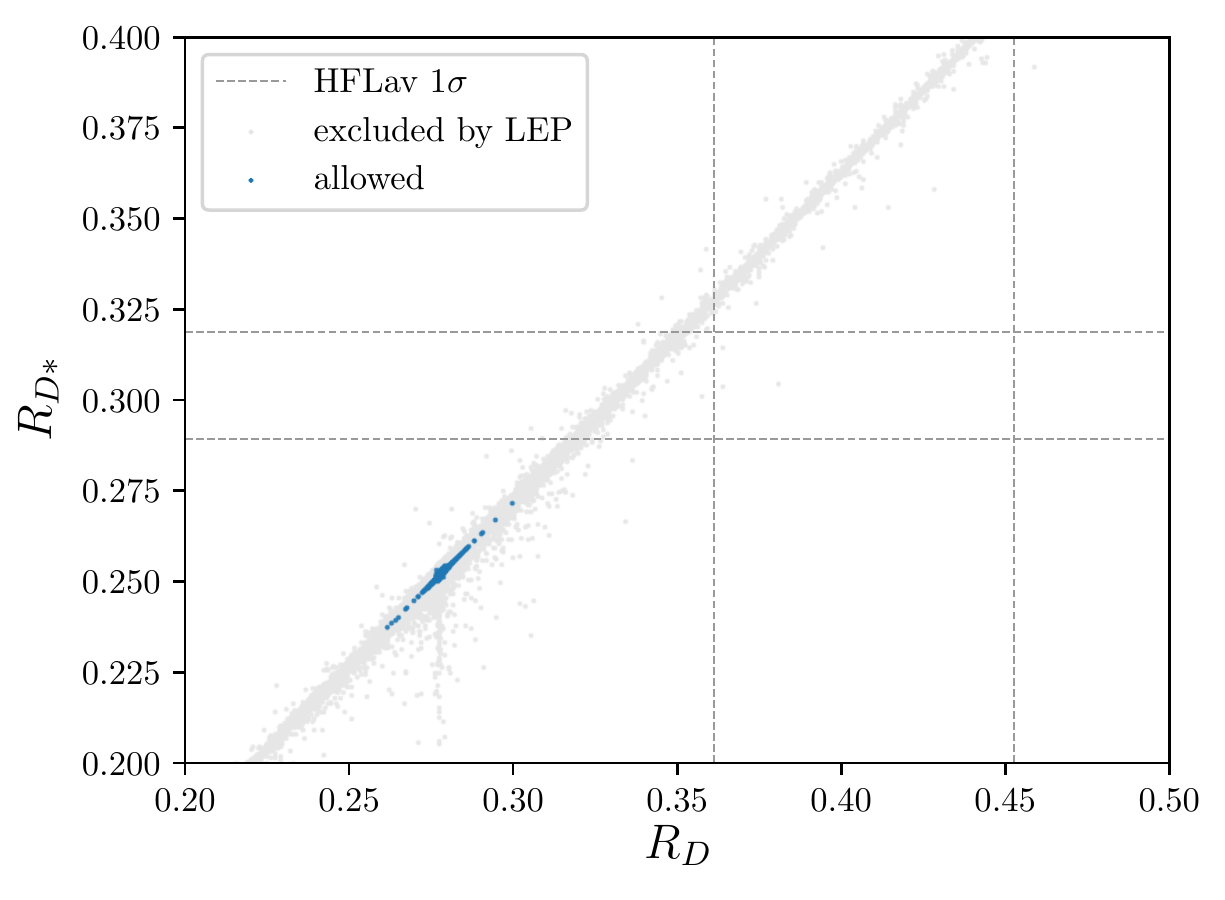}
\caption{Predictions for lepton flavor universality tests in $B\to D\tau\nu$ and $B\to D^*\tau\nu$ compared to the SM prediction and the experimental world
averages for all allowed points (dark blue) as well as for all points excluded by LEP $Z$ pole constraints (light gray).}
\label{fig:RD}
\end{figure}

Lepton-flavor universality in charged currents is also tested in the decays
$B\to D^{(\ast)}\tau\nu$ based on the $b\to c\tau\nu$ transition, that are experimentally
more challenging than the $B\to D^{(\ast)}\ell\nu$ decays with $\ell=e$ or $\mu$ that
are used to measure the CKM element $V_{cb}$.
In recent years, several measurements by BaBar, Belle, and LHCb 
\cite{Lees:2013uzd,Huschle:2015rga,Aaij:2015yra,Sato:2016svk,Hirose:2016wfn,Aaij:2017uff}
have consistently shown higher
values for the ratios
\begin{equation}
R_{D^{(\ast)}} = \frac{\Gamma(B\to D^{(\ast)}\tau\nu)}{\Gamma(B\to D^{(\ast)}\ell\nu)}
\end{equation}
than predicted, with small uncertainties, in the SM. A global combination by the
HFLAV collaboration finds a combined significance of around $4\sigma$ \cite{Amhis:2016xyh}.
In figure~\ref{fig:RD}, we show our predictions for $R_D$ and $R_{D^\ast}$ for all allowed points.
The dominant effects lead to a simultaneous increase (or decrease) of both ratios,
as observed by experiment, since they are generated by a vector operator with left-handed
quarks and leptons.
But although there are some points in parameter space where the tension with experiment can be
reduced compared to the SM, the overall size of the effects is too small to accommodate
the experimental central values.
The main reason for this is the limit on the size of the $\tau$ lepton fundamental Yukawa
coupling coming from $Z\to\tau\tau$ decays at LEP. Switching off the LEP constraints, we
find huge effects in both $R_D$ and $R_{D^\ast}$, as shown by the light gray points in figure~\ref{fig:RD}.
An interesting question is whether a non-minimal FPC model with a vanishing Wilson
coefficient for the operator $\mathcal{O}_{\Pi f}$ or some other protection of the $Z\tau\tau$ coupling exists that could accommodate
a large violation of LFU in $R_D$ and $R_{D^\ast}$. We leave the investigation 
of this question to a future analysis.

\subsubsection{Lepton flavor universality tests in FCNC decays}

\begin{figure}
\includegraphics[width=\textwidth]{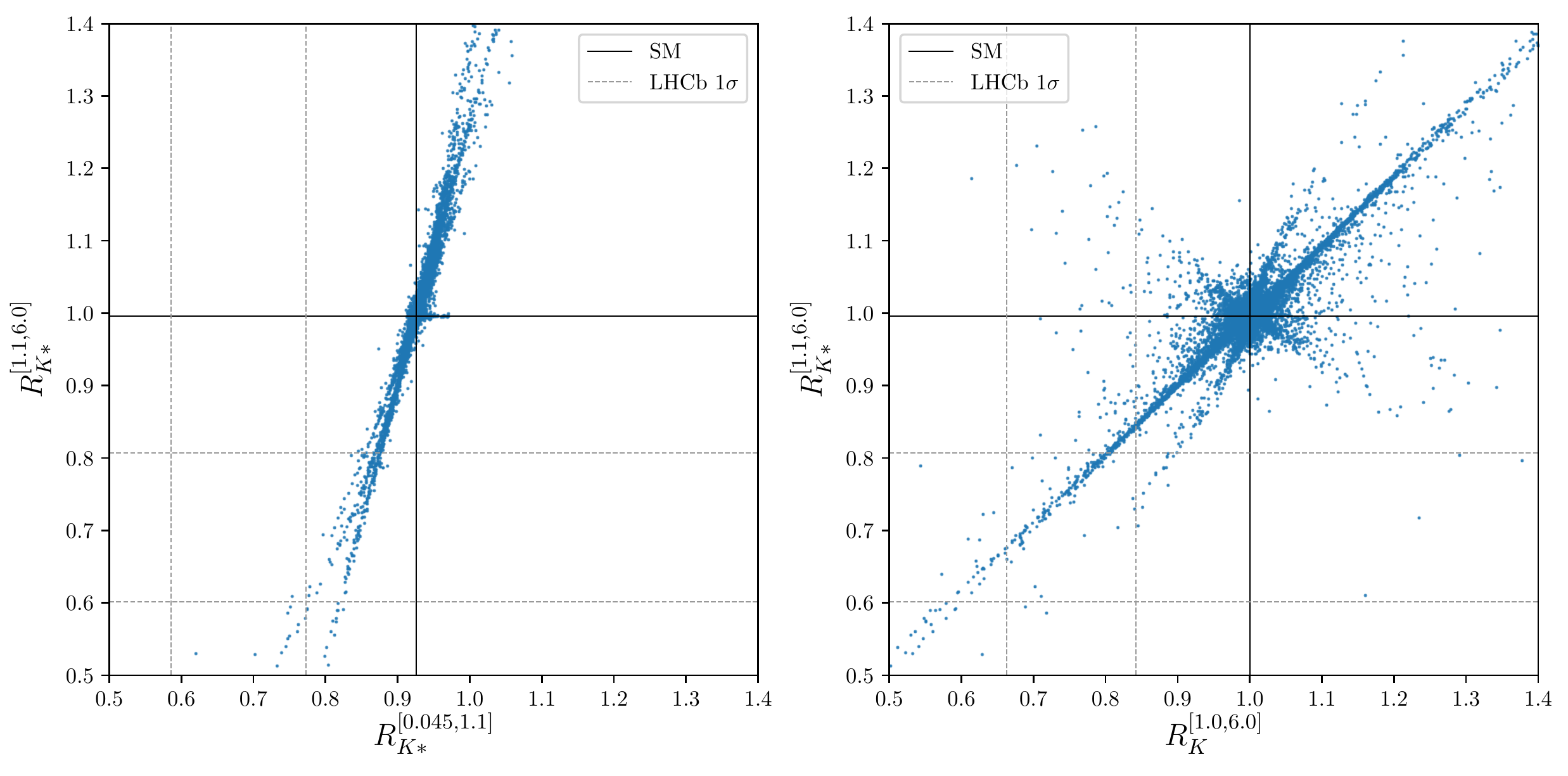}
\caption{Predictions for $\mu$-$e$ universality tests in $B\to K^*\ell^+\ell^-$ and $B\to K\ell^+\ell^-$ compared to the SM prediction and the LHCb measurements for all allowed points.}
\label{fig:RK}
\end{figure}

Measurements by the LHCb experiment of the ratios
\begin{equation}
R_{K^{(*)}}^{[a,b]} = \frac{\int_a^b dq^2 \, \frac{d\Gamma}{dq^2}(B\to K^{(*)}\mu^+\mu^-)}{\int_a^b  dq^2 \, \frac{d\Gamma}{dq^2}(B\to K^{(*)}e^+e^-)}
\end{equation}
show tensions with the theoretically very clean SM prediction at the level of 2--3$\sigma$
\cite{Aaij:2014ora,Aaij:2017vbb}. Several analyses have shown that these
tensions can be consistently explained by physics beyond the SM, in particular by a vector operator with left-handed quarks and muons \cite{Altmannshofer:2017yso,Capdevila:2017bsm,DAmico:2017mtc,Geng:2017svp,Ciuchini:2017mik,Hiller:2017bzc}.
As seen from the discussion in section~\ref{sec:fcnc}, such an operator is generated in MFPC as well, along with the analogous operator
with right-handed muons. In effective models of partial compositeness, it has been shown that the deviation in $R_{K^{(*)}}$ can be
explained if left-handed muons have a significant degree of compositeness\footnote{%
Alternative explanations with partial compositeness mostly using NP in the electronic channels have been suggested as well,
but cannot explain additional tensions present in  $b\to s\mu^+\mu^-$ transitions \cite{Carmona:2015ena,Carmona:2017fsn}.}
\cite{Niehoff:2015bfa} (see also \cite{Megias:2016bde,Megias:2017ove} for extra-dimensional constructions),
corresponding to a sizable fundamental Yukawa coupling in
MFPC. In figure~\ref{fig:RK}, we show our predictions for $R_{K}$ and $R_{K^{*}}$ for all allowed points in the bins measured by LHCb,
compared to the SM prediction and the experimental measurement. We find a significant number of points
where all three observations can be explained within 1--2$\sigma$, demonstrating that the MFPC model can explain all $R_{K^{(*)}}$
anomalies in terms of new physics. Since this comes about by means of an operator involving left-handed muons,
the model also fits the global fit to $b\to s\mu^+\mu^-$ observables, where additional tensions
are present (see e.g.\ \cite{Altmannshofer:2017fio}), much better than the SM.

\section{Conclusions}\label{sec:conclusions}

We have performed a comprehensive numerical analysis of flavor physics in Minimal Fundamental Partial Compositeness (FPC).
To the best of our knowledge, this is the first numerical analysis of a UV completion of partial compositeness
with a realistic flavor structure in the quark sector. Our main findings can be summarized as follows.
\begin{itemize}
\item Indirect CP violation in kaon mixing (measured by the parameter $\epsilon_K$)
is larger than observed in large parts of the parameter space, but we also find a large number
of points where it is small enough.
\item For the points allowed by the $\epsilon_K$ constraints, sizable effects in $B^0$ and $B_s$
mixing are observed for many points, including non-standard CP-violating mixing phases close to
the level currently probed in precision experiments.
\item While we impose the absence of charged lepton flavor violation for simplicity, the violation
of lepton flavor \textit{universality} (LFU) is unavoidable with partial compositeness. We find 
LFU tests like the ratios of $\pi$ or $K\to e\nu$ vs.\ $\mu\nu$ to constitute important constraints on
the parameter space.
\item LFU violation in $B\to D^{(*)}\tau\nu$, as currently indicated by several experiments at the level
of $4\sigma$, cannot be generated at a sufficient size to reproduce the experimental central values due
to LEP constraints on the $Z\tau\tau$ couplings. The tensions can however be ameliorated compared to the SM.
\item The MFPC model can explain both hints for LFU violation in $B\to Kee$ vs.\ $\mu\mu$ ($R_K$) and 
$B\to K^*ee$ vs.\ $\mu\mu$ ($R_{K^*}$) simultaneously, as shown in figure~\ref{fig:RK}.
\end{itemize}
To summarize, Minimal Fundamental Partial Compositeness is a predictive UV complete model with a
realistic flavor sector that can be tested by present and future flavor physics experiments. If the anomalies
in $R_D$ and $R_{D^*}$ are confirmed to be due to NP, a non-minimal model with protected $Z$ couplings
to tau leptons might be preferred. If the deviations in $R_K$ and $R_{K^*}$ are confirmed, they
could be first indications of technifermions and techniscalars coupling strongly to muons.

Our explorative study can be generalized in several ways.
There are additional low-energy precision tests that we have not considered, e.g.\  the anomalous
magnetic moment of the muon or electric dipole moments.
We have also not attempted to construct a realistic lepton sector explaining the origin of neutrino
masses or the absence of lepton flavor violation.
In contrast to effective models of partial compositeness, also the form factors of the new
strong interaction, that we have simply scanned here, could be computed in principle, boosting
the predictiveness of the model.

\subsection*{Acknowledgments}
PS would like to thank Christoph Niehoff for useful discussions.
The work of PS and DS was supported by the DFG cluster of excellence ``Origin and Structure of the Universe''. 
FS and AET acknowledge partial support from the Danish National Research Foundation grant DNRF:90.

\appendix
\section*{Appendix}

\section{Next-to-leading order operators for the kinetic terms} \label{app:NLOoperators}
Ref.~\cite{Cacciapaglia:2017cdi} listed all operators that modify the kinetic terms of the EW gauge bosons and the pNGBs at NLO. For completeness we refer here the operators which contribute to the EW precision parameters, $ S $ and $ T $. The leading operator contributing to the $ S $ parameter is 
	\begin{equation}
	\mathcal{O}_{WW} = \dfrac{1}{32 \pi^2} A^{I}_{\mu\nu} A^{J\mu\nu} \, \Tr \left[ T^{I}_\tcf \Sigma (T^{J}_\tcf)\transpose \Sigma^\dagger \right].
	\end{equation}
There are two kinds of operators contributing to the $ T $ parameter. Two operators are due to corrections from the EW gauge interactions,
	\begin{align}
	\mathcal{O}_{\Pi D}^1 &= \dfrac{1}{32} \dfrac{f_\TC^2}{16 \pi^2} \Tr \left[ (\Sigma \overleftrightarrow{D}_\mu \Sigma^\dagger) T_\tcf^{I} (\Sigma \overleftrightarrow{D}^\mu \Sigma^\dagger) T_\tcf^{I}  \right] \, ,\\
	\mathcal{O}_{\Pi D}^2 &= \dfrac{1}{32} \dfrac{f_\TC^2}{16 \pi^2} \Tr \left[ (\Sigma \overleftrightarrow{D}_\mu \Sigma^\dagger) T_\tcf^{I} \right] \Tr \left[ (\Sigma \overleftrightarrow{D}^\mu \Sigma^\dagger) T_\tcf^{I}  \right]\, ,
	\end{align}
and four operators are due to corrections from SM fermions,
	\begin{align}
	\mathcal{O}_{y\Pi D}^{1} &= \dfrac{1}{32} \, \dfrac{f_\TC^2}{(4 \pi)^4} (y_f^\ast y_f)^{a_1} \phantom{}_{a_2} \phantom{}^{i_1 i_2} (y_{f'}^\ast y_{f'})^{a_3} \phantom{}_{a_4} \phantom{}^{i_3 i_4}  (\Sigma^\dagger \overleftrightarrow{D}^\mu \Sigma)_{a_1} \phantom{}^{a_2} (\Sigma^\dagger \overleftrightarrow{D}^\mu \Sigma)_{a_3} \phantom{}^{a_4} \epsilon_{i_1 i_2} \epsilon_{i_3 i_4}\ , \\
	\mathcal{O}_{y\Pi D}^{3} &= \dfrac{1}{32} \, \dfrac{f_\TC^2}{(4 \pi)^4} (y_f^\ast y_f)^{a_1} \phantom{}_{a_2} \phantom{}^{i_1 i_2} (y_{f'}^\ast y_{f'})^{a_3} \phantom{}_{a_4} \phantom{}^{i_3 i_4}  (\Sigma^\dagger \overleftrightarrow{D}^\mu \Sigma)_{a_1} \phantom{}^{a_2} (\Sigma^\dagger \overleftrightarrow{D}^\mu \Sigma)_{a_3} \phantom{}^{a_4} \epsilon_{i_1 i_4} \epsilon_{i_2 i_3}\ , \\
	\mathcal{O}_{y\Pi D}^{4} &= \dfrac{1}{32} \, \dfrac{f_\TC^2}{(4 \pi)^4} (y_f^\ast y_f)^{a_1} \phantom{}_{a_2} \phantom{}^{i_1 i_2} (y_{f'}^\ast y_{f'})^{a_3} \phantom{}_{a_4} \phantom{}^{i_3 i_4}  (\Sigma^\dagger \overleftrightarrow{D}^\mu \Sigma)_{a_1} \phantom{}^{a_4} (\Sigma^\dagger \overleftrightarrow{D}^\mu \Sigma)_{a_3} \phantom{}^{a_2} \epsilon_{i_1 i_2} \epsilon_{i_3 i_4}\ , \\
	\mathcal{O}_{y\Pi D}^{6} &= \dfrac{1}{32} \, \dfrac{f_\TC^2}{(4 \pi)^4} (y_f^\ast y_f)^{a_1} \phantom{}_{a_2} \phantom{}^{i_1 i_2} (y_{f'}^\ast y_{f'})^{a_3} \phantom{}_{a_4} \phantom{}^{i_3 i_4}  (\Sigma^\dagger \overleftrightarrow{D}^\mu \Sigma)_{a_1} \phantom{}^{a_4} (\Sigma^\dagger \overleftrightarrow{D}^\mu \Sigma)_{a_3} \phantom{}^{a_2} \epsilon_{i_1 i_4} \epsilon_{i_2 i_3}\ .
	\end{align}
We have normalized these operators corresponding to the normalization of the decay constant in the LO kinetic terms, such that the corresponding strong coefficients are expected to be $ \mathcal{O}(1) $.

\bibliography{FlavorRefs.bib}

\end{document}